\documentclass[inte,nonblindrev]{./StyleFiles/informs3} 

\OneAndAHalfSpacedXII 

\usepackage{natbib}
\bibpunct[, ]{(}{)}{,}{a}{}{,}%
\def\bibfont{\small}%
\usepackage{multirow}
\usepackage[table,xcdraw]{xcolor}
\usepackage[noend]{algorithmic}
\usepackage{algorithm}
\usepackage{longtable,booktabs}
\usepackage{pdflscape,lipsum,afterpage,adjustbox,rotating}
\usepackage{comment}
\usepackage{bold-extra} 
\usepackage[font=small,labelfont=bf]{caption}
\usepackage{hyperref}
\hypersetup{
    colorlinks=true,
    linkcolor=blue,
    filecolor=blue,      
    urlcolor=blue,
    citecolor= blue,
}

\newcommand{\iind}{n}
\newcommand{\jind}{q}
\newcommand{\kind}{r}
\newcommand{\mind}{s}

\newcommand{\Iset}{\mathcal{I}}
\newcommand{\Jset}{\mathcal{J}}
\newcommand{\Kset}{\mathcal{K}}
\newcommand{\Mset}{\mathcal{M}}
\newcommand{\Lset}{\mathcal{L}}
\newcommand{\Eset}{\mathcal{E}}

\newcommand{\ShiftSwaps}{Swaps}

\begin{document}



\RUNTITLE{Nurse Scheduling at Toronto's Long-Term Care Homes}

\TITLE{Optimization Helps Scheduling Nursing Staff at the Long-Term Care Homes of the City of Toronto}

\ARTICLEAUTHORS{%
\AUTHOR{Manion Anderson, Merve Bodur, Scott Rathwell, Vahid Sarhangian}
\AFF{Department of Mechanical \& Industrial Engineering, University of Toronto, Toronto, ON, CANADA \EMAIL{\{manion.anderson,scott.rathwell\}@mail.utoronto.ca, \{bodur,sarhangian\}@mie.utoronto.ca} \URL{}}
} 

\ABSTRACT{%
The City of Toronto Long Term Care Homes \& Services (LTCH\&S) division is one of the largest providers of long-term care in the Canadian province of Ontario, providing care to 2,640 residents at 10 homes across Toronto. Our collaboration with LTCH\&S was initiated to facilitate the increasingly challenging task of scheduling nursing staff and reduce high absenteeism rate observed among the part-time nurses. We developed a spreadsheet-based scheduling tool to automate the generation of schedules and incorporate nurses’ preferences for different shifts into the schedules. At the core of the scheduling tool is a hierarchical optimization model that generates a feasible schedule with the highest total preference score while satisfying the maximum possible demand. Feasible schedules had to abide by a set of complex seniority requirements which prioritized more senior nurses when allocating the available shifts. Our scheduling tool was implemented in a 391-bed home in Toronto. The tool allowed nursing managers to generate feasible schedules within a fraction of an hour, in contrast to the status-quo manual approach which could took up to tens of hours. In addition, the schedules successfully accounted for preferences with on average above 94\% of the allocated shifts ranked as most preferred. 
}%



\maketitle
\section{Introduction}\label{intro}
{\bf Background and Motivation.} The growing proportion of seniors in the population has led to a steady increase in demand for long-term care services in North America. By 2050, the U.S. is expected to have 19 million elderly persons in need of some type of long-term care, more than double the 9 million who required long-term care services in the year 2000 \citep{spetztrupinbatescoffman2015}. As of 2016, there were 15,600 nursing homes operating 1,660,400 long-term care beds throughout the country. The trend is similar in Canada. In 2018, the province of Ontario spent over C\$4.28 billion on long-term care, accounting for more than 7\% of the total health sector expenses for the province \citep{ontario.ca_2019}. With a rapidly growing demand, managing nursing resources for long-term homes has also become increasingly challenging. In 2018, a survey conducted by the Ontario Long Term Care Association found that 80\% of care homes across the province had difficulty filling nursing shifts and 90\% of care homes experienced difficulty recruiting new nursing staff \citep{TILTC2019}. As such, there is significant interest in improving the management of nursing resources with the goal of reducing costs and improving the quality of care.


The City of Toronto Long Term Care Homes \& Services (LTCH\&S) Division is one of the largest providers of long-term care in Ontario, providing care to 2,640 residents at 10 separate homes across Toronto. The division provides 24-hour, resident-focused care and service, from fundamental services such as personal care, medical care, and housekeeping, to specialized services such as community outreach, language and cultural partnerships, and short-stay respite beds. In order to provide consistent, high-quality care to its residents, the division employs over 3,400 staff and schedules over 573,000 shifts each year. Over two-thirds of all staff employed by the division work within the nursing department, responsible for providing nursing and personal care within the homes. The workforce is comprised of Registered Nurses (RN), Registered Practical Nurses (RPN), and Personal Support Workers (PSW). 

Our collaboration with the division was initiated to address two main challenges faced by the division with respect to scheduling the required nursing staff in the long-term care homes - in particular part-time nurses. (1) Status-quo practice involved manually constructing and revising feasible schedules by nursing managers of each home. Full-time nurses were assigned to fixed rosters, but the part-time schedules needed to be constructed in each planning horizon depending on their availability and the remaining demand to be filled. Due to the relatively large number of staff to be scheduled and the complex scheduling requirements (which we elaborate on further below), constructing and revising schedules required a considerable effort and consumed a significant amount of time. Time which would ideally be spent on improving care for the residents. (2) The division faced a high rate of absenteeism, in particular among its part-time nursing staff. Over the course of 2016, more than 12.5\% of assigned shifts were not worked as scheduled due to call-ins, sick leave, vacation, and other forms of absenteeism. The status-quo scheduling practice did not account for nursing preferences. As a result, part-time nurses faced high-uncertainty with respect to the number and timing of the shifts they would be allocated. Our collaborators at the LTCH\&S division believed that this uncertainty contributed to the high absenteeism rate. 

 
\textbf{Summary of the approach and contributions.} We developed a spreadsheet-based scheduling tool to automate the process of generating schedules. The primary component of the tool is a hierarchical optimization model. The model generates a feasible schedule that maximizes the satisfied demand for each scheduling unit across a planning horizon of 6 weeks. The model takes the (deterministic) demand over the horizon and the schedule for full-time nurses as inputs, and assigns the available part-time nurses to the remaining shifts after allocating the full-time nurses. In addition, in order to reduce absenteeism for part-time staff, the model incorporates the nurses' shift preferences. (The preferences are collected from the part-time staff prior to each scheduling cycle.)  More specifically, among the schedules that achieve the maximum demand, the model returns the one(s) with maximum overall preference of the assigned nursing staff.

In developing the model and solution approach we addressed a number of technical and practical challenges, which enabled the implementation of our scheduling tool in practice. In particular, in addition to typical nurse scheduling constraints, the schedules had to abide by the so called \emph{Armstrong seniority rules}, according to which more senior nurses (i.e., who have worked with the City for a longer time) receive higher priority when allocating the demand. Although the motivation behind the seniority rules were known to the division managers, there was a lack of consensus with regards to certain details, in particular how the rules were to be interpreted when in conflict with other scheduling requirements. The interpretation varied from home to home and even within a home in some cases. As such, we developed a precise definition of the seniority rules through interviews with nurse managers and scheduling clerks, and incorporated them into our optimization model as constraints. Indeed, accounting for the seniority of staff is common when scheduling workforce and is not unique to the City of Toronto. The seniority rules which we formalize in this work, provide a general framework for prioritizing the allocation of demand to staff with different seniority levels and extend the previous models in the literature (see the discussion below on related literature). 

To ensure the adaptation of the tool in practice, we opted to develop an Excel-based user interface which the scheduling staff was already familiar with. In addition, to develop an inexpensive proof of concept, we decided to use an open source solver (OpenSolver). However, due to the complexity of the optimization model, in particular the large number of constraints, the required time to generate an optimal schedule in OpenSolver became prohibitively large. As such, we also developed an approximate version of the model which together with a heuristic algorithm can generate optimal or near-optimal schedules using OpenSolver and within a reasonable time. We illustrate the quality of the generated schedules by comparing them with the optimal ones obtained using a commercial solver (Gurobi) for problem instances based on real data.

Finally, we implemented the scheduling tool in 3 pilots at a 391-bed home in Toronto. For each pilot, we used the tool to generate a six-week schedule for two units comprised of a total of 83-90 nurses with different specialties. In order to execute the pilots, we worked with the division to collect nursing preferences and inform nurses on the value of incorporating their preferences into the scheduling procedure. Through these subsequent pilots we improved the scheduling tool and addressed several transitional challenges. The pilots demonstrated the significant reduction in the amount of time spent on developing schedules (from multiple days to a fraction of a hour). In addition, the generated schedules successfully incorporated preferences, with all except one scheduling pool having above $92\%$ of the allocated shifts ranked with the highest preference score.

\textbf{Related literature.} There is a large body of literature on the nurse-rostering (or scheduling) problem, mainly focusing on scheduling nurses in clinical units of hospitals; see \cite{burke_2004a} and \cite{decausmaecker_2011} for comprehensive literature reviews. Despite the large amount of literature related to nurse scheduling, \cite{burke_2004a} note that the majority of work is not implemented in practice nor utilizes real data. \cite{Kellogg_2007} find that only about 30\% of the models proposed in the literature are actually implemented. The authors indicate the lack of early communication  with the nurses and those in charge of scheduling, lack of nurse-centered solutions, and the limited relationships between the academics and third party scheduling software vendors as the major contributors. In this work, we addressed such challenges by ensuring that the nurses and nurse scheduler were involved at all steps of the project from formulating the problem to providing feedback on the schedules created; by including nursing preferences for shifts into the problem formulation; and by developing a spreadsheet-based software solution which is easy to use and does not require the involvement of a third-party vendor. See \cite{ovchinnikov2008spreadsheet} for another example of spreadsheet-based scheduling software. 

Recognizing the need for better nursing management in long-term care homes, a number of recent studies have focused on optimizing task scheduling, skill-mixes, and staffing levels. For instance, \cite{lieder_2015} present a dynamic optimization approach to minimize the earliness and tardiness of executed care tasks from the residents' preferred times, and \cite{bekker_2019} present a stochastic optimization approach to develop new shift patterns which improve the waiting times and service levels based upon fluctuating care demand throughout the day. \cite{slaugh2018consistent} and \cite{slaugh2020positional} investigate staffing strategies to improve continuity (or consistency) of care. In contrast, in this work we focus on optimal scheduling of part-time nurses while taking their preferences into account, in order to reduce absenteeism.

An integral component of our model is the set of seniority rules which all schedules must abide by. The Armstrong seniority rules are unique to the City of Toronto, but share similarities with other seniority rules proposed and studied in the literature. \cite{caron_1999} introduce seniority constraints into the classical assignment problem. They consider two types of seniority constraints; weak and strong. The weak constraints are satisfied when an unassigned person can only be scheduled by removing a scheduled shift from a person of the same or higher seniority. The strong seniority constraints are satisfied under the same condition but must hold for a chain of reassignments rather than individual reassignments. An individual reassignment occurs when a single shift is assigned from one nurse to another. A chain of reassignments consists of a series of reassignments (not necessarily associated with the same shift). \cite{volgenant_2004} discusses the application of the assignment problem with seniority constraints to scheduling float nurses at a hospital. \cite{topaloglu_2009} studies the problem of scheduling medical residents at a hospital unit. The residents are grouped into different seniority levels based on their experience. Residents' working preferences are taken into account using a multi-objective optimization model which uses hard constraints to enforce seniority-based work rules and soft constraints to prioritize seniority when allocating workload and assigning shifts based on days of requests. 

The inclusion of seniority rules into our model differs from previous work along two dimensions. First, previous studies assume full availability of the workforce for the available shifts. In contrast, in our model each nurse is available for a subset of shifts and hence availability and seniority need to be considered jointly when assigning nurses to shifts. In addition, the seniority rules considered in our work are more general than those previously considered in the literature. Under Armstrong's Seniority rules, different seniority tiers are assigned a minimum shift requirement. Once the minimum shift requirements are met, the seniority rules become similar to the seniority constraints discussed by \cite{caron_1999}. The difference is that a more senior nurse must have the same or more shifts than a less senior nurse, relative to the minimum shift requirements of their corresponding tiers. The inclusion of minimum shift requirements considerably increases the complexity of the model as the seniority must be prioritized when assigning shifts until the minimum shift requirements are met. This means a higher seniority employee should always receive their minimum shift requirement before an employee of lower seniority receives even one shift. 

Different types of preferences have been incorporated into the nurse-scheduling problem in the literature. \cite{bellanti_2004} include preferences for which days/shifts to work on and which days/shifts to take off, \cite{downsland_2000} introduce preferences for different shift patterns and sequences, and \cite{bard_2005} assign a penalty cost depending on the degree to which individual preferences of nurses are violated. We incorporated preferences by requiring nurses to provide a numeric score to indicate their low or high preference for all shifts which they are available for. We consider the total preference score of a feasible schedule as a measure of how favorable the schedule is and as a proxy for absenteeism. Other modeling approaches for incorporating nurse preferences include using soft constraints (\citealt{berrada_1996}) or as an additional objective within a multi-objective optimization approach (\citealt{bard_2005}).

Here, we only account for preferences after making sure the maximum possible demand is satisfied. We do so using a hierarchical model that first finds the maximum possible demand that can be satisfied while adhering to all the constraints, and then finds a schedule with maximum total preference score which achieves the maximum demand and again satisfies all the constraints. Other two-phase models have been proposed in the literature for the nurse scheduling problem, but for different purposes. Examples include obtaining feasible schedules in the first phase and subsequently improving them in the second (\citealt{warner_1976}), and scheduling specific portions of the schedule in the first phase (e.g., days off, vacations, or rotations) and then constructing the actual nurse schedules in the second (\citealt{valouxis_2012}).

\textbf{Organization of the rest of the paper.} The remainder of the paper is organized as follows. In Section \ref{ProbDesc}, we provide a detailed description of our nurse scheduling problem including the objectives and the requirements that the schedules must meet. In Section \ref{SolApp}, we outline our proposed solution approaches, with the detailed description of the optimization models and the heuristic algorithm provided in the Appendix. Section \ref{Sec:UserInterface} provides an overview of the user interface of our scheduling tool and provides an example of a schedule output. In Section \ref{ImpRes}, we describe the implementation of the scheduling tool in three pilots and present the results. Finally, we provide conclusions and a discussion of future directions in Section \ref{Conc}.

\section{Problem Description}\label{ProbDesc} 

The nursing department within each home at the LTCH\&S division is split into units. Each unit employs nurses with three designations; RNs, PRNs, and PSWs. Nurses in each of the designations have a different skill set. Although RNs are in principle qualified to work as RPNs or PSWs, this is to be avoided due to concerns regarding continuity of care. As such, we do not consider scheduling across designations. 
 
Schedules are developed at the unit level and for each designation, which we refer to as a \emph{scheduling pool} for the remainder of the paper. Further, schedules are developed for a six-week horizon, referred to as a \emph{cycle}, which contains three two-week pay periods or \emph{blocks}.

Each RN, RPN, and PSW is assigned to a given scheduling pool as either a full-time or part-time nurse. Full-time nurses are assigned a rotating schedule which repeats every six weeks, guaranteeing the same shifts in each cycle as mandated by a collective bargaining agreement. As such, we focus on scheduling part-time nurses to fill remaining demand. The demand for each designation in each unit remains constant in time, but varies among the morning (7AM to 3PM), evening (3PM to 11PM), and night (11PM to 7AM) shifts. The demand for part-time nurses within a scheduling pool for a specific shift is therefore determined by subtracting the total number of full-time nurses scheduled for that shift, from the total demand for the shift. Although the total demand is determined by the number of beds in each unit and remains constant in time, the demand for part-time nurses in a unit may still vary due to vacations and leave requests of full-time nurses. 

Previously, within each cycle part-time nurses were required to provide their availability for different shifts, without the option of specifying a higher preference for some shifts over the others. Our proposed approach requires part-time nurses to also include a preference score for each shift they are available for. The scores take values in $\{0,1,2,3\}$ with 0 indicating unavailability, and a higher non-zero value indicating a \emph{higher} preference. We define the \emph{total preference score} of a schedule as the sum of the preference scores of the scheduled part-time nurses for their assigned shifts.
 
The scheduling problem can be described as follows. Given the demand over the cycle; find a feasible schedule (i.e., an assignment of the part-time nurses to each unit of demand) for each scheduling pool and for each shift, that conforms with the scheduling requirements, and satisfies the maximum possible demand. In addition, among feasible solutions that satisfy the maximum possible demand, select the (possibly non-unique) solution which has the maximum total preference score. In the following, we describe these hierarchical objectives as well the scheduling requirements in detail. 
 
\subsection{Objectives}
We consider two possibly conflicting objectives: (1) maximizing the satisfied demand and (2) maximizing the total preference score. The process of filling the unfilled demand is cumbersome, costly, and sometimes unsuccessful. Unfilled demand results in additional workload for other nurses and could potentially impact the quality of care. The unfilled demand is filled either through communication with existing nurses to see if they are able to extend their availability, work overtime, or by scheduling nurses across units and hence possibly compromising continuity of care. Therefore, we opt to prioritize demand satisfaction over preference maximization. That is, we aim to select the schedule with the maximum total preference score among those satisfying the maximum possible demand. In doing so, we do not explicitly account for the uncertainty in the total satisfied demand due to absenteeism and its dependence on the provided preferences. (See Section \ref{Conc} for further discussion.)

\subsection{Solution Requirements}
 
We divide the requirements into two categories; general scheduling requirements and seniority requirements.

\textbf{General scheduling requirements.} The following requirements are mandated by the collective bargaining agreement, and the division scheduling policies. All schedules must satisfy these requirements.
\begin{itemize}

\item \emph{Availability}: a nurse can only be scheduled for shifts he/she is available for.

\item  \emph{Maximum number of assigned shifts}: a nurse cannot be assigned overtime in any given two-week block within a six week cycle. This means that a maximum of 10 shifts can be assigned to each nurse within a two-week block. 

\item  \emph{No back-to-back shifts}: a nurse cannot be scheduled for a shift within 11-hours of another scheduled shift. Since all shifts have the same length of 8 hours with fixed start and end times, this rule implies that any pair of shifts must be scheduled a minimum of two shifts apart. 

\item  \emph{Maximum Number of Weekend Shifts}: a nurse cannot be scheduled for more than ten shifts on a weekend (Saturdays and Sundays) within a six-week cycle.

\item  \emph{No overbooking}: a nurse cannot be scheduled for a shift if the demand for that shift has been filled.
\end{itemize}

\textbf{Armstrong seniority requirements}. The Armstrong Seniority rules (which we formalize in this work) prescribe a method for allocating demand to part-time nurses based upon their seniority level within their scheduling pool. Seniority rankings for each nurse are re-calculated after every scheduling cycle and can vary from cycle to cycle as nurses accumulate working hours, or as new nurses are added and/or existing nurses are removed from a unit. The following elements are critical in understanding the requirements.

\begin{figure}[htbp] 
\centering
\includegraphics[scale = 0.70]{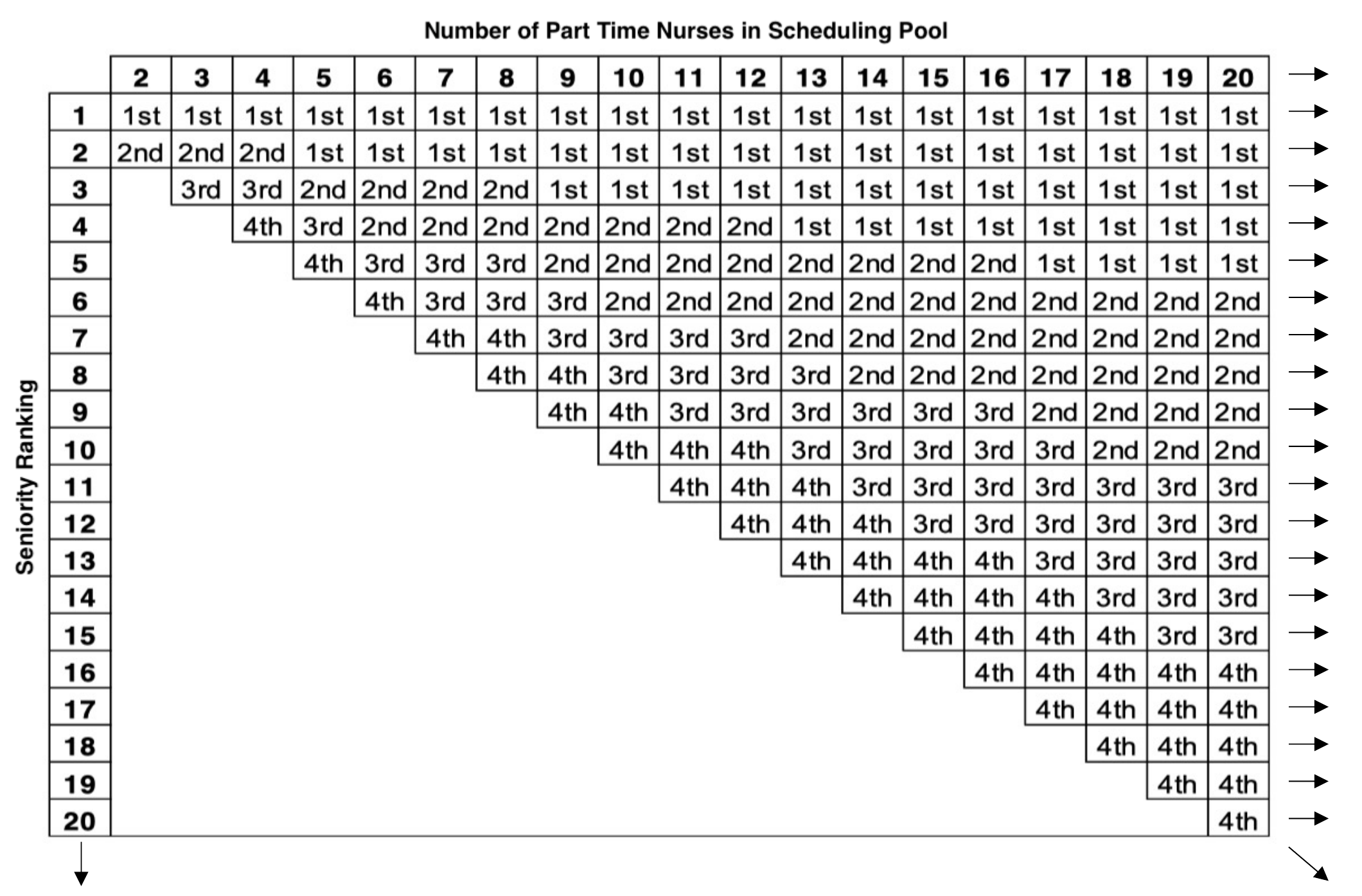}
\caption{The Armstrong tier allocation chart: each nurse is assigned to a seniority rank based on the number of nurses available in the pool (columns) and their seniority ranking within the pool (rows).}
\label{ArmstrongTiers}
\end{figure}

\emph{Armstrong tiers}: Nurses in each scheduling pool are divided into four tiers using the tier allocation chart presented in Figure \ref{ArmstrongTiers}. The assignment is based on the total number of part-time nurses in the scheduling pool and their seniority rankings within their pool. 

\emph{Minimum shift requirement}: In order to generate a schedule that abides by the Armstrong Seniority rules, each nurse must be assigned a minimum number of shifts. The minimum shift requirement is the same for all nurses of the same tier and corresponds to the number of shifts that a nurse should receive within each of the three blocks of a cycle as long as enough demand is available. We denote the minimum shift requirement for nurse $i\in\{1,\ldots,n\}$ in block $k\in\{1,2,3\}$ by $g_{ik}$, where $n$ is the total number of part-time nurses available.   

\emph{Armstrong Delta}: The Armstrong Delta, defined for each scheduling block and each nurse in the scheduling pool, is the total number of shifts assigned to the nurse in each block subtracted by their minimum shift requirement. If the Armstrong Delta is exactly 0 for a nurse in a given block this means the nurse has been assigned exactly their minimum shift requirement. If the Armstrong Delta is negative the nurse has been assigned less shifts than their minimum shift requirement, and if the Armstrong Delta is positive they have been assigned more shifts than their minimum shift requirement. We denote the Armstrong Delta for nurse $i$ in block $k$ by $\delta_{ik}$.

\emph{Rules}: The following set of rules must hold in order for a schedule to adhere to the Armstrong Seniority rules.
\begin{itemize}
    \item {\it Rule 1A: Prioritize seniority for shift allocation until minimum shift requirements are met.} The first part of Rule 1 requires that a nurse in a lower seniority tier should not be assigned any shifts if a more senior nurse has not yet been assigned their minimum shift requirement unless an exception is occurred. (We will elaborate on the exceptions below.) Formally,
    \begin{equation}
    \delta_{ik} < 0 \implies  \delta_{i'k} \leq - g_{i'k},  \quad \forall i \in \{1,..,n-1\}, i' = \{i+1,\hdots,n\} , k \in \{1,2,3\}, 
    \end{equation}
    that is, if nurse $i$ has a negative $\delta_{ik}$, then the 
    Armstrong delta
    of all nurses in lower seniority tiers should be equal to the negative value of their minimum shift requirement, i.e., they should not be allocated any shifts. 

    \item {\it Rule 1B: Prioritize seniority for shift allocation until minimum shift requirements are met.}
    The second part of Rule 1 states that a nurse with higher seniority should not receive more than their minimum shift requirement if any of the less senior nurses have not yet received their minimum shifts requirement, unless an exception has occurred.
    Formally,
    \begin{equation}
      \delta_{i'k} < 0 \implies  \delta_{ik} \leq 0,  \quad \forall i \in \{1,..,n-1\}, i' = \{i+1,\hdots,n\} , k \in \{1,2,3\}, 
    \end{equation}
    that is, if a nurse has a negative $\delta_{ik}$, then the 
    Armstrong delta of all nurses with higher seniority should not be greater than 0.

    \item {\it Rule 2: Prioritize sequential shift allocation once minimum shift requirements are met.}
    The second rule ensures that shifts are added sequentially and in the order of seniority, once all nurses have received their minimum shift requirements (unless an exception is occurred). 
    Formally,
    \begin{equation}
    \delta_{ik}, \delta_{i'k} \geq 0 \implies (\delta_{i'k} + 1) \geq \delta_{ik} \geq \delta_{i'k}, \quad \forall i \in \{1,..,n-1\}, i' = \{i+1,\hdots,n\} , k \in \{1,2,3\},
    \end{equation}
    that is, if a pair of nurses both have 
    Armstrong delta values greater than 0, the more senior nurse should have a $\delta_{ik}$ equal to or at most one greater than the less senior nurse.
\end{itemize}

\emph{Exceptions}: 
An exception allows an Armstrong seniority requirement or requirements to be relaxed in order to satisfy the general scheduling requirements. For example, assume that in a given block, a nurse with the highest seniority ranking is available for one less shift than his/her minimum shift requirement. If Rule 1A were to be enforced strictly, none of the nurses in the scheduling pool of lower seniority would be assigned to even a single shift, even if there was unfilled demand. Instead, the requirement is relaxed since the nurse cannot be scheduled for any more shifts without breaking the \textit{availability} general scheduling requirement. Similarly, an Armstrong seniority rule may be relaxed for a given nurse if they have been assigned their \textit{maximum number of assigned shifts} for the block; if they have been assigned their \textit{maximum number of weekend shifts} for the cycle; if a scheduled shift will result in \textit{back to back} shifts; or if a scheduled shift will result in \textit{overbooking}. An example detailing how the Armstrong Seniority rules and exceptions can be checked using the output of our scheduling tool is presented in Section \ref{Sec:UserInterface}.

\section{Solution Approaches}\label{SolApp}
In this section, we provide an overview of the two developed solution approaches and relegate the details to the Appendix. Both approaches involve sequentially solving two Integer Programs (IPs) - one for demand maximization and the other for preference maximization - but differ in terms of the optimality of the solution and required computational time. The implemented solution approach employs an approximate version of the Armstrong seniority rules and allocates shifts up to the minimum requirements of the seniority tiers. The remaining demand, if any, is then allocated using a heuristic algorithm. While this approach may lead to a sub-optimal schedule, it can generate a feasible schedule in a reasonable amount of time using Open Solver in Excel, even when the number of nurses in the pool is ``large". In the second approach, we model all seniority requirements in an exact manner. This approach guarantees the optimality of the constructed schedule, but does not scale well as the number of nurses increases. We use the results of the exact model in Section \ref{ImpRes} to demonstrate the quality of the schedules generated by the implemented approach.

\begin{figure}[htbp] 
\centering
\includegraphics[scale = 0.615]{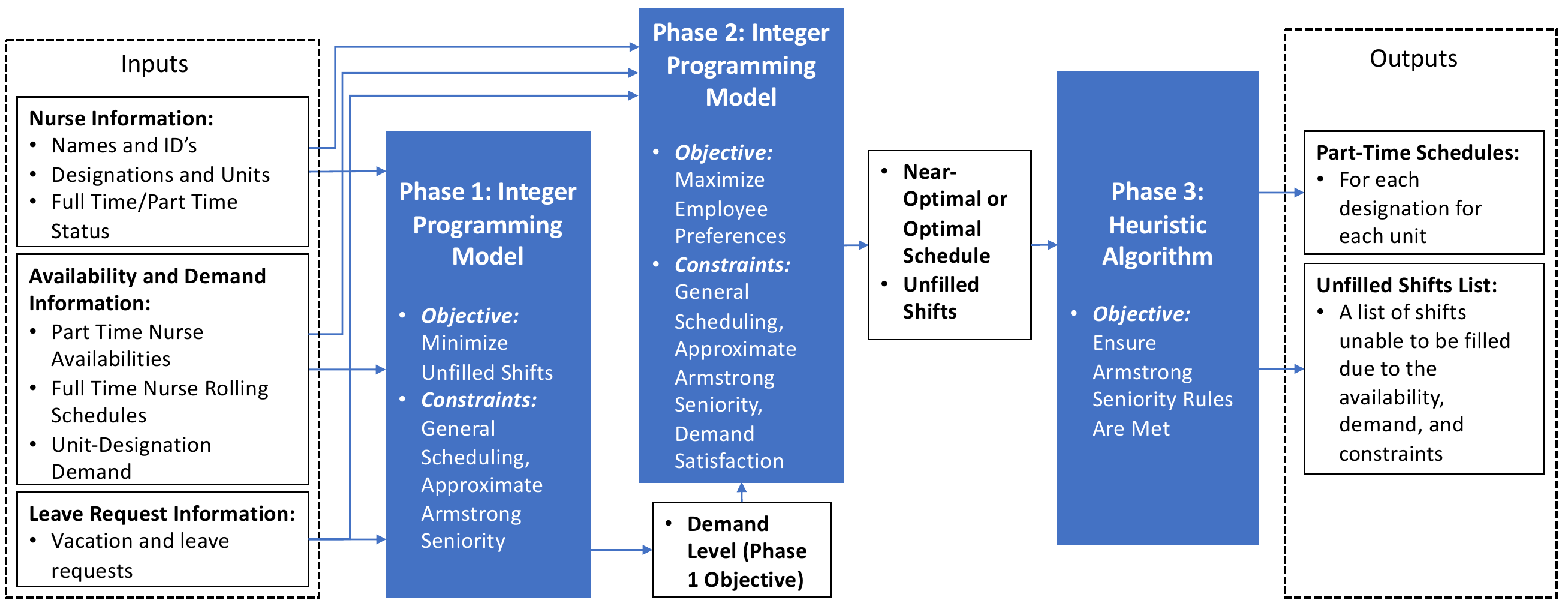}
\caption{Context diagram of the implemented solution approach}
\label{ContextDiagram}
\end{figure}

Figure \ref{ContextDiagram} presents a context diagram of the implemented solution approach including the inputs and outputs of each phase. In the following we provide details of each phase (Sections \ref{Phase12} and \ref{Phase3}) as well the exact solution approach (Section \ref{Exact}).

\subsection{{\textbf{Integer Programming Models (Phase 1 and 2)}}}\label{Phase12}
The main decision variables for both phases are binary variables indicating whether each nurse in the scheduling pool is assigned to any of the shifts within the scheduling horizon. The objective in Phase 1 is to maximize the satisfied demand over the horizon subject to the general scheduling and seniority requirements. The objective in Phase 2 is to maximize the total preference score of all the assignments. In addition to the general scheduling and seniority requirement constraints, an additional constraint is included to ensure that the maximum (feasible) demand obtained in Phase 1 is achieved. That is, Phase 2 finds a schedule which maximizes the preferences while satisfying the maximum possible demand. 

In the implemented approach, the constraints enforce an approximate version of the Armstrong seniority rules. The approximate constraints become exact when the total demand for a scheduling pool is less than the sum of all minimum shift requirements for the nurses in the scheduling pool (as long as no exceptions occur). Although this is often the case for scheduling pools at the LTCH\&S division, the total demand can exceed the total minimum requirements. If this scenario occurs, the implemented (approximate) IP has a constraint which ensures that a nurse can only be allocated up to the same number of shifts as their minimum shift requirement and no more. Any remaining shifts are then allocated using a heuristic algorithm which we discuss next. For details on the constraint sets of the IPs see Appendix \ref{Apendix:MIPModel}.

\subsection{\textbf{Heuristic Algorithm} (Phase 3)}\label{Phase3}
The schedule generated in Phase 2 is used as a starting solution for the heuristic algorithm to allocate the remaining demand while adhering to the constraints. Here we provide an overview of the idea. The full algorithm as well as a more detailed description can be found in Appendix \ref{Apendix:Heuristic}. 

The algorithm starts by greedily assigning any unfilled demand to the most senior nurse who is available for the shift and can be scheduled without violating any of the general scheduling requirements. (If no nurse can be assigned to the shift, it will be added to the list of unfilled demand.) Next, it goes through all assigned shifts and checks whether each shift can be reassigned to another nurse -- who is currently not assigned to that shift but is eligible to be assigned -- without violating any of the Armstrong seniority rules.  Once this procedure is executed for all assigned shifts in the planning horizon, a single iteration of the algorithm is complete. The algorithm is then iterated multiple times until no reassignments are possible. 

The algorithm terminates after a finite number of iterations (in our experiments at most three iterations were sufficient) and the output schedule is guaranteed to meet the general scheduling and Armstrong seniority rules, although it may be sub-optimal in terms of the unfilled demand and/or preferences of the nurses. (See Appendix \ref{Apendix:Heuristic} for details and  formal arguments). In Section \ref{ImpRes}, we conduct numerical experiments using real data to demonstrate that the heuristic performs quite well compared to the exact IP model which we describe next.


\subsection{\textbf{Exact Integer Programming Model}}\label{Exact}
\label{subsec:ExactIP}

The exact IP model has a similar structure to the implemented one, except that the Armstrong seniority rules are exactly enforced. To this end, we developed a set of exact constraints which utilize fifteen binary variables to ensure that all rules and exceptions are enforced simultaneously. These binary variables are used to indicate whether a nurse has hit an exception or not for a given block based upon the shift demand and all assigned shifts over the scheduling cycle. This enables Armstrong rules to be relaxed if an exception occurs, so that a nurse is not preventing other nurses from receiving additional shifts. The exact formulation provides an optimal feasible schedule for any combination of demand and availability. The full model and further description can be found in Appendix \ref{Apendix:MIPModel}. Although this approach provides optimal results, the binary variables are activated using a large number of logical constraints utilizing ``big-M" constraints. These constraints are known to make IP models challenging to solve. In Section \ref{ImpRes} we use the exact formulation to illustrate the performance of the implemented approach for ``smaller" instances of the problem.


\section{User Interface} \label{Sec:UserInterface}
\begin{figure}[htbp]
\centering
\includegraphics[scale = 0.65]{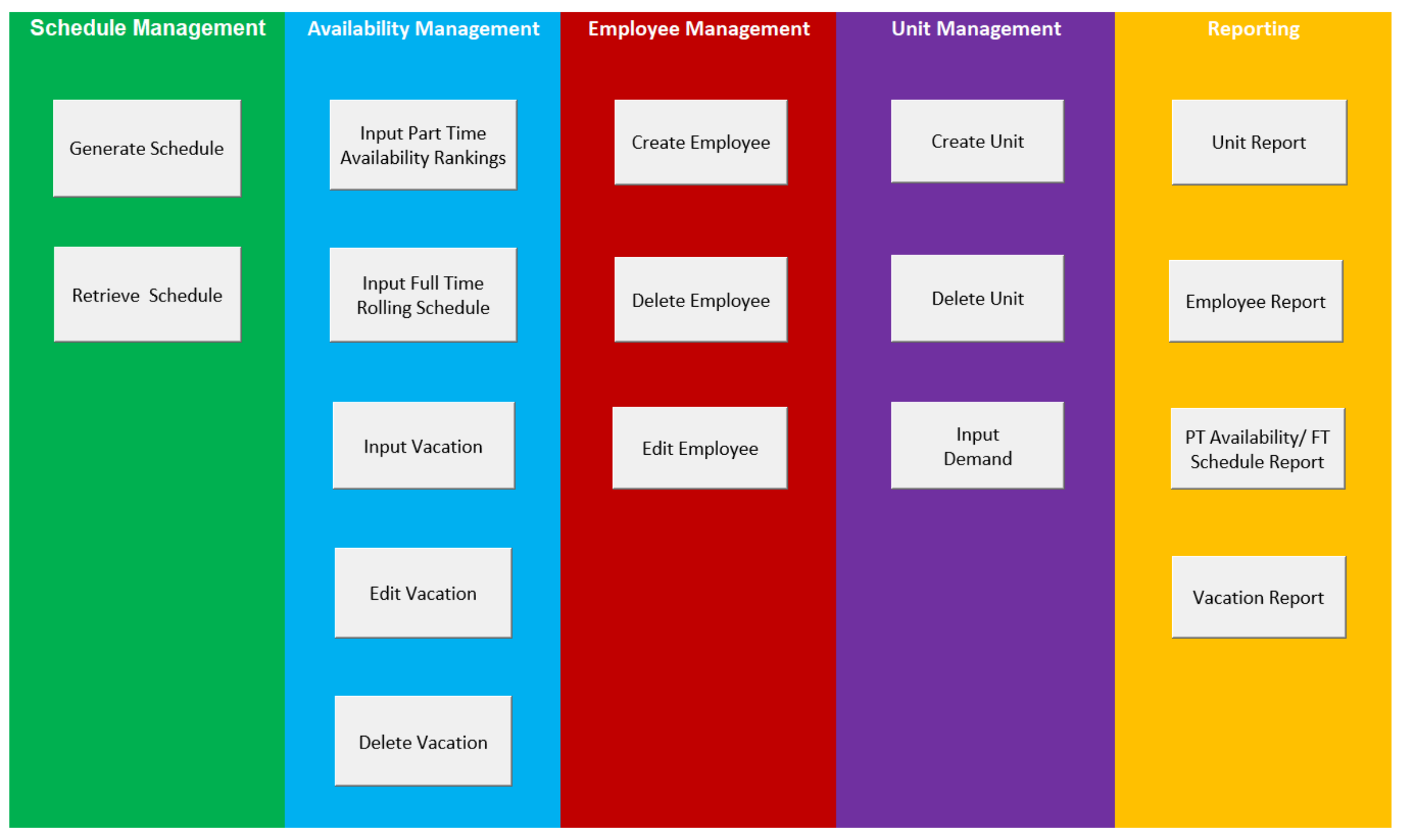}
\caption{The user interface of the scheduling tool}
\label{fig:UserInterface}
\end{figure}

We developed the user interface in Excel using a combination of macros and user forms. The interface consists of five modules; Unit Management, Employee Management, Availability Management, Schedule Management, and Reporting. Each of these modules includes a set of functions which enable the user to input data, edit previously recorded data, generate a schedule, or generate reports. A screenshot of the Excel interface with functions for each module is depicted in Figure \ref{fig:UserInterface}. 
The Unit, Employee, and Availability Management modules are all used to input data required for solving the IPs. Within the Unit Management module, the Create Unit function is used to create a database for each scheduling pool within a unit. Within the Unit Management module, the Create Employee function is used to create employee records and assign them to a scheduling pool. The Edit Employee function is used to edit an employee's designation, full-time/part-time status, scheduling pool assignment, and seniority level. Within the Availability Management module, the functions are used to input availability and preferences of part-time nurses, as well as the rolling schedule and vacation requests of full-time nurses. 

The Reporting module is used to view and reconcile data between the Excel tool and a legacy LTCH\&S employee management system. This includes making sure demand levels, nurse seniority, availability, full-time schedules, and vacation requests are all up to date. The schedule management module is used to generate part-time nurse schedules.

The Generate Schedule function can be used to generate a schedule for all scheduling pools, scheduling pools within a specific unit, scheduling pools of a specific designation, or for a specific scheduling pool. In order to generate a schedule, a scheduling cycle must be selected and an optional time limit can be provided. This time limit is for the solve time for each of the IPs in the model. If the time limit is hit and a feasible solution has not been found then no solution is provided. Once the schedule is generated, a macro in Excel outputs the schedule in a readable format. This consists of printing the schedule in three separate blocks each with the part-time demand, unfilled demand, and average preference score of nurses assigned to each shift. The output also includes a summary of the preference scores for each nurse, including the percentage of  assigned shifts that were most preferred, preferred, and least preferred. Finally, the output displays an unassigned shift code for each shift in which a nurse was available, but was not assigned to that shift as well as a summary of the minimum shift requirements and the total number of shifts assigned to each nurse in each block. Together, these outputs can be used to check that a schedule abides by the general scheduling and Armstrong seniority requirements. A summary of the unassigned shift codes and the corresponding explanations are provided in Table \ref{tab:ExceptionCodes}. 

\begin{table}[htbp]
\caption{Description of unassigned shift codes}
\label{tab:ExceptionCodes}
\footnotesize
\centering
\begin{tabular}{|c|l|}
\hline
\textbf{Unassigned Shift Code} & \textbf{Description}                                                                                                                                 \\ \hline
B                       & \begin{tabular}[c]{@{}l@{}}Cannot be scheduled or else back-to-back constraint \\ is violated (Two shifts within 11 hours of each other)\end{tabular}    \\ \hline
D                       & \begin{tabular}[c]{@{}l@{}}Demand has been filled by either more or less \\ senior nurse(s) according to the Armstrong seniority rule\end{tabular}  \\ \hline
M                       & \begin{tabular}[c]{@{}l@{}}The nurse has already received their maximum \\ of 10 shifts within the two-week block\end{tabular}                        \\ \hline
W                       & \begin{tabular}[c]{@{}l@{}}The nurse has already been scheduled for the maximum \\ of 10 weekend shifts in the six-week planning horizon\end{tabular} \\ \hline
\end{tabular}
\end{table}

\begin{figure}[htbp] 
\centering
\includegraphics[scale = 1.2]{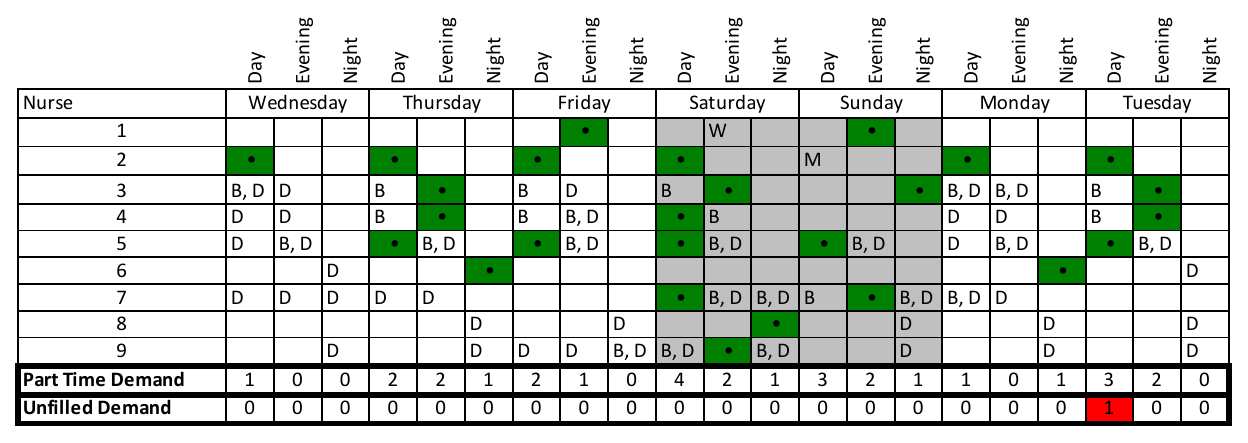}
\caption{Partial output of the scheduling tool for one week of a planning horizon}
\label{fig:ExampleSchedule}
\end{figure}

\textbf{Example}. Figure \ref{fig:ExampleSchedule} presents an example of a generated schedule for one week of a six-week planning horizon. The full schedule including other summary measures can be found in Appendix \ref{Apendix:ScheduleOutput}. The output shown includes the assigned shifts (marked in green) to each nurse, unassigned shifts (marked with codes presented in Table \ref{tab:ExceptionCodes}), part-time demand for each shift, and the total unfilled demand in each shift (marked in red). The cells corresponding to weekends are shaded gray to help nursing managers check for weekend requirements. The total number of shifts assigned to each nurse, minimum shift requirements, and Armstrong deltas is also generated as a part of the schedule output and can be found in Table \ref{tab:ExampleSchedTotals}. Next, we will detail how the schedule in Figure \ref{fig:ExampleSchedule} together with the summary values in Table \ref{tab:ExampleSchedTotals} can be utilized by the nursing managers to verify that the seniority requirements are satisfied. 

\begin{table}[htbp]
\caption{Total number of assigned shifts and Armstrong deltas for the schedule presented in Figure \ref{fig:ExampleSchedule}}
\label{tab:ExampleSchedTotals}
\footnotesize
\centering
\begin{tabular}{|l|c|c|c|}
\hline
\textbf{Nurse \#} & \textbf{\begin{tabular}[c]{@{}l@{}}Min. Shift \\ Req.\end{tabular}} & \textbf{\begin{tabular}[c]{@{}l@{}}Shifts Assigned\\ (Block 3)\end{tabular}} & \textbf{\begin{tabular}[c]{@{}l@{}}Armstrong Delta\\ (Block 3)\end{tabular}} \\ \hline

 1           & 8                        & 4            & -4           \\ \hline
 2           & 8                        & 10           & 2            \\ \hline
 3           & 8                        & 8            & 0            \\ \hline
 4           & 6                        & 7            & 1            \\ \hline
 5           & 6                        & 7            & 1            \\ \hline
 6           & 4                        & 5            & 1            \\ \hline
 7           & 4                        & 5            & 1            \\ \hline
 8           & 3                        & 4            & 1            \\ \hline
 9           & 3                        & 4            & 1            \\ \hline
\end{tabular}
\end{table}

When examining the summary values in Table \ref{tab:ExampleSchedTotals}, at first it appears that Armstrong Rule 1A is violated for the third block of the schedule. The potential violation is due to Nurse 1 having an Armstrong delta of zero and all nurses of lower seniority having non-negative Armstrong deltas. We can however verify that an exception has occurred. Observe that the nurse has an unassigned shift code ``W" for Saturday evening shifts in both the first and second week of the third block. This implies that the nurse could not be scheduled for any more shifts in the block due to the \textit{maximum number of weekend shifts} general scheduling requirement. 

It also appears that Armstrong Rule 2 is in violation for the third block of the schedule. The potential violation is due to the third nurse having an Armstrong delta of zero while the second nurse has an Armstrong delta of two and the fourth through ninth nurses have  Armstrong deltas equal to one. It can again be verified that an exception has occurred. In the schedule, all shifts that the nurse was available for but remained unfilled have either a ``B" or ``D" or both unassigned shift codes. The ``B" shift code signifies that the shift could not be scheduled without breaking the \textit{no back-to-back shifts} general scheduling requirement. The ``D" shift code signifies that either no demand exists for the shift or that the demand was filled by either more or less senior nurses according to the Armstrong seniority requirements. As long as no shifts marked with ``D" can be reassigned to the third nurse from nurse two or nurses four through nine, the third nurse has hit an exception. Observe from Figure \ref{fig:ExampleSchedule} that the third nurse has the shift code ``D" for the Wednesday evening and Friday evening shifts but neither nurse two nor nurses four through nine are scheduled for either of those shifts. Therefore, the second nurse has hit an exception and Rule 2 is not violated.

\section{Implementation and Results}\label{ImpRes} 
In this section we discuss the execution of a series of pilots executed at a large care home operated by the City of Toronto LTCH\&S division. We also present the challenges faced during implementation and detail the quantitative and qualitative results achieved. 

\subsection{Pilot Setting and Timeline}
We conducted three pilots at a 391-bed home within the City of Toronto LTCH\&S division. Two out of the seven units within the home, consisting of a total of roughly 90 full- and part-time nurses were included in the pilots. Each pilot required a total of six schedules to be generated for three scheduling pools (corresponding to different specialties) within each of the two units. The size of the scheduling pools ranged from 3 to 12 nurses, which is representative of the size of scheduling pools in most homes across the city. Schedules were required to be generated three weeks in advance of the six-week scheduling cycle in order for the schedules to be posted two weeks in advance as per the division's policy. Table \ref{tab:PilotDates} presents the timeline of the pilots. 

\begin{table}[htbp]
\caption{Timeline of the pilots}
\label{tab:PilotDates}
\footnotesize
\centering
\begin{tabular}{|l|l|l|l|}
\hline
\textbf{Pilot} & \textbf{Employees} & \textbf{Scheduling Cycle} & \textbf{Schedule Post Date} \\ \hline
1 & 91 & \begin{tabular}[c]{@{}l@{}}December 18th/2019 - \\ January 28th/2020\end{tabular} & December 4th/2019              \\ \hline
2 & 92 & \begin{tabular}[c]{@{}l@{}}January 29th/2020 - \\ March 10th/2020\end{tabular}     & January 15th/2020              \\ \hline
3 & 83    & \begin{tabular}[c]{@{}l@{}}March 11th/2020 - \\ April 14th/2020\end{tabular}       & February 26th/2020             \\ \hline
\end{tabular}
\end{table}

\subsection{Data}
We collected the required data for the pilots from multiple sources. The required data included nursing demand, full-time nurse schedules, part-time nurse availability, and leave requests. The demand and full-time nurse schedule data were obtained from the division's legacy scheduling system. The legacy scheduling system ensures all human resources related data is automatically up to date and is utilized to store availability and leave data and manage schedule changes and call-in shifts. The availability of part-time nurses during the targeted cycles as well as leave requests had to be collected directly from the nurses. A process for collecting leave requests was already in place and consequently no additional efforts were required. While there was also an existing process for collecting availability, preferences were not collected. Therefore, we worked with the division to develop a new form and process for collecting preferences. The process of collecting preferences began roughly seven weeks prior to the beginning of the scheduling cycle. The process consisted of a project lead distributing the new form to all part-time nurses in the units. Throughout the weeks leading up to the first distribution of the new ranked availability forms, we helped the division to conduct information sessions explaining the purpose of the pilot and providing instructions on how to fill the new forms. The forms were then collected by the project lead three weeks prior to the start of the scheduling cycle and manually entered into our scheduling tool.

\subsection{Challenges of the Implementation}
The implementation of the new scheduling tool involved coordination with various stakeholders, revising existing procedures, and dealing with unforeseen technical challenges. When we initially began to collect ranked availability forms from the nurses, we observed that a majority of shifts were ranked as most preferred. This may have been due to concerns that assigning low preference scores to shifts would lead to a smaller number of shifts eventually allocated. As such, with support from the nursing managers and in the subsequent information sessions we clarified to the nurses that the rankings will not affect the number of allocated shifts, and encouraged them to provide truthful preferences.

Another challenge was the poor synchronization of the processes that were used to collect and input data required by the model. The late submission of ranked availability forms in the first pilot forced nurse managers to use historical availability data without rankings for some nurses. Similarly, late submission of leave requests forced nurse managers to make manual changes to the schedules generated by the model. We determined new timelines for the subsequent pilots and improved the User Interface functions to facilitate the process of data entry. The latter was particularly challenging since the schedules had to be posted two weeks prior to each cycle, leaving us little time to address last-minute errors in the User Interface or questions around the generated schedule.

Finally, as alluded to earlier, formally defining the Armstrong seniority rules and convincing the stakeholders that the generated schedules complied with them was particularly challenging. As we began to formally define the Armstrong seniority rules using existing documentation, as well as interviews with nurse managers, nurses, and scheduling clerks, we uncovered a lack of consensus with regards to the exceptions where seniority rules could be relaxed and whether general scheduling or Armstrong rules were to be prioritized. We worked with all stakeholders to develop a formal definition of the seniority rules. The output generated by the User Interface which simplified the process of verifying seniority rules was instrumental in convincing the division that the generated schedules comply with the seniority rules, and in turn implementation of the tool in practice.

\subsection{Quantitative Results}
\label{subsec:QuantitativeResults}
In order to test the quality of solutions generated by the implemented (approximate) approach, we compared its results with those of the exact IP in terms of running time and sub-optimality of both objectives. We utilized data from the first two pilots in our experiments, including real preference and availability data collected from part-time nurses. As mentioned above, the number of nurses to be scheduled in each pool ranged from 3 to 12. To examine how the solution time scales for larger pools we further combined data from multiple scheduling pools to create four larger units (ranging from 19 to 22 employees). 

The results for the 15 scheduling pools are summarized in Table \ref{tab:RunTimes}. For each pool and solution approach, the table presents the total part-time demand (same for all solution approaches), total unfilled demand in the generated schedule, the percentage of total shifts assigned that were most preferred, and the time it took to obtain the schedule (Run Time). Pools 1-11 are based on real availability and demand data, whereas  pools 12-15 correspond to synthetic cases constructed by combining multiple scheduling pools. The results for the exact model are obtained using both Open Solver in Excel and Gurobi in Python.

\begin{table}[htbp]
\caption{Performance summary of the generated schedules for 15 scheduling pools using the implemented approach and the exact approach. TL indicates that the solver was not able to find an integer solution before the 1 hour time limit per phase (not including setup time in Excel). Run times for each solution approach are reported in seconds.}
\label{tab:RunTimes}%
\footnotesize
  \centering
    \begin{tabular}{cccccccccccc}\toprule
    & & \multicolumn{3}{c}{Implemented Model (OpenSolver)}            & \multicolumn{3}{c}{Exact Model (OpenSolver)}            & \multicolumn{3}{c}{Exact Model (Gurobi)} \\
    \cmidrule(r){3-5}
    \cmidrule(r){6-8}
    \cmidrule(r){9-11}
      \multicolumn{1}{c}{Pool} & \multicolumn{1}{c}{\begin{tabular}[c]{@{}l@{}}\# of PT \\ Nurses\end{tabular}} &\multicolumn{1}{c}{\begin{tabular}[c]{@{}l@{}}Unfilled Demand \\ (Total Demand) \end{tabular}} & \multicolumn{1}{c}{\begin{tabular}[c]{@{}l@{}} \% First \\ Preference\end{tabular}} & \multicolumn{1}{c}{\begin{tabular}[c]{@{}l@{}}Run \\ Time \end{tabular}} &\multicolumn{1}{c}{\begin{tabular}[c]{@{}l@{}}Unfilled \\ Demand\end{tabular}} & \multicolumn{1}{c}{\begin{tabular}[c]{@{}l@{}}\% First \\ Preference\end{tabular}} & \multicolumn{1}{c}{\begin{tabular}[c]{@{}l@{}}Run \\ Time \end{tabular}} &\multicolumn{1}{c}{\begin{tabular}[c]{@{}l@{}}Unfilled \\ Demand\end{tabular}} & \multicolumn{1}{c}{\begin{tabular}[c]{@{}l@{}}\% First \\ Preference\end{tabular}} & \multicolumn{1}{c}{\begin{tabular}[c]{@{}l@{}}Run \\ Time \end{tabular}} \\
          \midrule
    1  & 3  & 0 (31)   & 100\%  & 42   & 0 & 100\% & 152   & 0   & 100\% & 1 \\
    2  & 5  & 0 (36)   & 100\%  & 63   & 0 & 100\% & 652   & 0   & 100\% & 1 \\
    3  & 5  & 2 (37)   & 54\%   & 98   & 2 & 54\%  & 668   & 2   & 54\%  & 1 \\
    4  & 8  & 0 (72)   & 100\%  & 201  & 0 & 100\% & 9356  & 0   & 100\% & 9 \\
    5  & 9  & 16 (162) & 94\%   & 243  & - & -     & TL    & 12  & 92\%  & 107 \\
    6  & 10 & 0 (79)   & 100\%  & 485  & - & -     & TL    & 0   & 100\% & 13 \\
    7  & 10 & 0 (85)   & 98\%   & 291  & - & -     & TL    & 0   & 98\%  & 12 \\
    8  & 11 & 0 (156)  & 100\%  & 474  & - & -     & TL    & 0   & 100\% & 2542 \\
    9  & 11 & 1 (168)  & 93\%   & 344  & - & -     & TL    & 1   & 97\%  & 105 \\
    10 & 11 & 0 (84)   & 92\%   & 329  & 0 & 93\%  & 50222 & 0   & 93\%  & 6.9 \\
    11 & 12 & 5 (163)  & 100\%  & 470  & - & -     & TL    & 5   & 100\% & 342 \\
    12 & 19 & 0 (240)  & 98\%   & 5765 & - & -     & TL.   & 208 & 100\% & TL \\
    13 & 20 & 0 (246)  & 97\%   & 1174 & - & -     & TL    & 208 & 100\% & TL \\
    14 & 21 & 0 (218)  & 98\%   & 7307 & - & -     & TL    & 196 & 100\% & TL \\
    15 & 22 & 0 (248)  & 99\%   & 1487 & - & -     & TL    & 243 & 100\% & TL \\ 
\bottomrule
    \end{tabular}%
\end{table}%


We observe that the implemented solution approach generates optimal or close to optimal schedules both in terms of the number of unfilled demand and the preference scores. In 8 out of the 15 cases (Pools 1,2,3,4,6,7,8,11) the implemented approach provides the optimal solution in terms of both unfilled demand and shift preferences. The percentage of first preferences for Pool 3 is lower compared to other pools (54\%) since the model prioritizes satisfying the demand. We note however that preferences are still taken into account with the rest of assigned shifts all having a preference score equal to two. 

In two out of the remaining 7 non-optimal cases (Pools 8 and 9) the implemented approach obtains a schedule with the optimal total unfilled demand but with a slightly lower percentage of first preference shifts. There is only one case (Pool 5) in which the implemented model provides a schedule with both a larger unfilled demand and a less favorable preference score. Pool 5 corresponds to a special instance where the number of available nurses was significantly lower than the total demand. Hence, a considerable number of shifts were allocated using the heuristic, leading to a sub-optimal solution. This scenario is an uncommon one, since the division follows strict staffing ratios to ensure that the majority of demand can be met using the available staff. We observe that the proposed approach generates optimal or close to optimal solutions when the demand does not significantly exceed the available supply of part-time nurses. 

The implemented approach performs very well for the remaining synthetic pools (12,13,14,15), with all pools having no unfilled demand and a large percentage of first preference shifts assigned (above $97\%$). For these larger pools, the exact model did not generate feasible solutions within the time limit using OpenSolver and generated feasible but sub-optimal solutions using Gurobi. The solutions provided by Gurobi had more than 85\% of the total demand unfilled, compared to the implemented approach which provided schedules with zero unfilled demand. 

As the pool size increases, the number of constraints increases exponentially and hence run times for the exact model could be very long, even with the commercial solver. In contrast, the running times for the implemented solution approach scales well as the problem size increases, and the solutions are optimal or close to optimal (for cases where the optimal solution is known) especially when demand is less than the total minimum shift requirements. This ability to generate high-quality schedules in a reasonable time was imperative to the implementation of our scheduling tool in practice. 

\subsection{Qualitative Results}
We were able to hear first hand from nurse managers, scheduling clerks, and the home administrators about how the scheduling tool helped to improve the scheduling process. A major improvement was the significant reduction in the amount of time required to generate and revise a schedule. A process which could take up to tens of hours over the course of two weeks, was now completed within 15 minutes for the majority of schedules. This subsequently allowed the nurse manager to allocate their attention to other tasks including improving the quality of care for the residents of the home. Compared to manually constructed schedules which could be sub-optimal (with respect to demand satisfaction) or do not meet some of the requirements, the generated schedules were also guaranteed to adhere to all scheduling requirements and satisfy the maximum possible demand. 

As illustrated in the previous section, the schedules achieved a high preference score, with the majority of nurses receiving their most preferred shifts. Since absenteeism data was not systematically collected by the division, we were unable to conduct a formal empirical evaluation of the impact of incorporating preferences into the schedules on the absenteeism rate. However, we expect that assigning part-time nurses to their preferred shifts would reduce the absenteeism rate and hence reduce costs and efforts required to fill unfilled demand.

\section{Conclusion}\label{Conc} 
We developed and implemented a spreadsheet-based software to automate and optimize the generation of schedules for nursing staff at long-term care homes of the LTCH\&S division of the City of Toronto. An important component of the optimization model is a set of seniority constraints, which (when possible) guarantees a minimum number of shifts for each seniority tier and prioritizes more senior staff (in receiving shifts) after allocating the minimum requirements. Explicit modeling of the seniority rules was crucial in making sure the tool can be implemented. In addition, formalization of the seniority rules and incorporating them into the tool has allowed the division managers to easily examine their impact on the number of shifts assigned to less-senior staff.

Upon successful completion of the pilots, we initiated discussions around the expansion of the tool to other homes. As the COVID-19 pandemic began in March 2020, major changes were made to the staffing structure, including temporary status change of part-time nurses to full time. As such, our plans were temporarily delayed. Indeed, the pandemic and its impact on long-term care homes has further demonstrated the need for novel and advanced approaches to managing nursing staff.

Upon city-wide implementation of the tool, we aim to empirically measure the potential reduction of absenteeism rate. This would require collecting pre- and post-implementation data on the absenteeism rate among part-time staff and conducting a statistical comparison of the pre- and post-implementation rates. The data can be further used to better understand and estimate the relationship between the provided preferences as well as other shift or nurse-related factors (see, e.g., \citealt{mudaly2015factors}) on the probability of a shift remaining unfilled. The estimated probabilities can then be directly incorporated into the optimization model to maximize the total expected satisfied demand, e.g., through a stochastic programming approach. 

Besides extensions to the model, integrating the spreadsheet tool with other information systems to automate the process of inputting the required data into the scheduling tool would significantly facilitate its use. We plan to incorporate these features into the future versions of the tool.

\bibliographystyle{StyleFiles/informs2014}
\bibliography{reference}



%
%
%

\newpage
\begin{APPENDICES}

\section{Integer Programming Model}
\label{Apendix:MIPModel}
The proposed model 
aims to find a schedule that maximizes nurse preferences while satisfying the maximum possible demand, i.e., shifts, given nurses' availability. The two interpretable outputs of the model are the \emph{schedule} and any \emph{unfilled demand} (that will need to be filled and managed outside of the model). A few practical assumptions (that are not limiting from the modeling perspective) are:

\begin{itemize}
  \item The model does not incur overtime. This shall be managed outside of the model.
  \item Full-time staff already have a set schedule, thus, only part-time scheduling will be addressed.
  \item Nurse $i$ has higher seniority than nurse $i+1$.
  \item The first day of each week is Wednesday.
  \item Schedules are built for a 6-week \emph{horizon} which is made up of 3 two-week scheduling \emph{blocks}.
  \item Scheduled shifts are all of 8-hour length.
\end{itemize}

In what follows, we provide objectives functions and constraints used in the model, for which the notation (indices, parameters, and decision variables) is given in Table \ref{t:notation}.


\begin{center}
\begin{longtable}{ll}
\caption{Nomenclature}
\label{t:notation}
\\
\toprule
\textbf{\textsc{Indices:}} \\*[0.1cm] 
$i$ & Employee index \\
$j$ & Shift index \\
$k$ & Block index  \\
$m$ & Weekend index \\
$e$ & Exception index \\
\midrule
\midrule
\textbf{\textsc{Sets:}} \\*[0.1cm]
$\Iset$ & Employee indices, $\{ 1,\hdots,\iind \}$ \\
$\Jset$ & Shift indices, $\{ 1,\hdots,\jind \}$ \\
$\Kset$ & Block indices, $\{ 1,\hdots,\kind \}$ \\
$\Mset$  & Weekend indices, $\{ 1,\hdots,\mind \}$ \\
$\Lset_m$ & 
Indices of shifts that belong to weekend $m$ \\ 
$\Eset$ & Exception types, $\{ available, maxout, backtoback, weekend, demand \}$ \\
\midrule
\midrule
\multicolumn{2}{l}{\textbf{\textsc{Parameters:}}} \\*[0.1cm]
$\iind$ & Number of employees in the scheduling unit \\
$\jind$ & Number of shifts in each two-week scheduling block \\
$\kind$ & Number of two-week scheduling blocks in 6-week scheduling horizon \\
$\mind$ & Number of weekend days in each block \\
$d_{jk}$ & Demand for shift $j$ in scheduling block $k$ after adjusting for 
pre-scheduled shifts \\
$g_{ik}$ & Number of shifts guaranteed to employee $i$ in scheduling block $k$ based on Armstrong rule \\
$g^{max}$ & Maximum number of shifts a employee can be assigned per block \\ 
$w$ & Minimum number of required weekend days off in scheduling horizon per employee
\\
$y_{ijk}$ & 1 if employee $i$ is available on shift $j$ in scheduling block $k$; 0 otherwise \\
$r_{ijk}$ & Rank of shift $j$ in scheduling block $k$ for employee $i$, where \\
& 0 = unavailable, and $1, \hdots,3$ = most to least preferred 
\\*[0.05cm] 
\midrule
\midrule
\multicolumn{2}{l}{\textbf{\textsc{Decision variables:}}} \\*[0.1cm]
\multicolumn{2}{l}{\footnotesize \textbf{For General Scheduling Constraints:}} \\*[0.1cm]
\multicolumn{2}{l}{\footnotesize \textit{Main Scheduling Variables:}} \\*[0.1cm]
$X_{ijk}$ & 1, if employee $i$ is scheduled on shift $j$ in scheduling block $k$; 0 otherwise \\
$S_{jk}$  & Amount of unfilled demand for shift $j$ in scheduling block $k$ \\
$D^*$ & Maximum demand that can be filled over the entire scheduling horizon \\*[0.2cm]
\multicolumn{2}{l}{\footnotesize \textit{Exception Related Variables:}} \\*[0.2cm]
$F_{ijk}^{available}$ & 1 if employee $i$ can be assigned shift $j$ as they are available; 0 otherwise \\*[0.14cm]
$F_{ijk}^{maxout}$ & 1 if employee $i$ can be assigned shift $j$ without exceeding 10 shifts in block $k$; 0 otherwise \\*[0.14cm]
$F_{ijk}^{backtoback}$ & 1 if employee $i$ can be scheduled for shift $j$ without being scheduled within two shifts of \\
& another scheduled shift or without already being scheduled for that shift; 0 otherwise \\
$F_{ijk}^{weekend}$ & 1 if employee $i$ can be scheduled for shift $j$ without exceeding 10 weekend shifts across all \\
& three blocks in the scheduling cycle; 0 otherwise \\
$F_{ijk}^{demand}$ & 1 if employee $i$ can be scheduled for shift $j$ without breaking an Armstrong seniority rule, \\
& i.e., if the demand of the shift has not been filled by more senior employees; 0 otherwise \\*[0.10cm]
$F_{ijk}$ & 1 if $F_{ijk}^{available}, F_{ijk}^{maxout}, F_{ijk}^{backtoback}, F_{ijk}^{weekend}, F_{ijkp}^{demand} $ are all equal to 1; 0 otherwise \\*[0.1cm]
$M_{ik}$ & 1 if employee $i$ is unable to be scheduled for any more shifts in block $k$; 0 otherwise \\*[0.15cm]
\multicolumn{2}{l}{\footnotesize \textbf{For Approximate Armstrong Constraint Set:}} \\*[0.1cm]
$\sigma_{ik}$ & 1 if employee $i$ is assigned their minimum shift requirement or an exception has occurred \\
& ($M_{ik}$ = 1) in scheduling block $k$; 0 otherwise \\
\multicolumn{2}{l}{\footnotesize \textbf{For Approximate and Exact Armstrong Constraint Sets:}} \\*[0.1cm]
$\theta_{ik}$ & 1 if employee $i$ has met or exceeded their minimum shift requirement in block $k$; \\ 
& 0 otherwise \\
\multicolumn{2}{l}{\footnotesize \textbf{For Exact Armstrong Constraint Set:}} \\*[0.1cm]
$\delta_{ik}$ & The difference between the minimum shift requirement and the number of shifts assigned \\
& for employee $i$ in block $k$ \\ 
$\pi_{ik}$ & 1 if employee $i$ has exceeded their minimum shift requirement in block $k$ ($\delta_{ik}$ $\geq$ 1); \\
& 0 otherwise \\
$\alpha_{ik}$ & 0 If employee $i$ has not met their minimum shift requirement and is available to be scheduled \\
& for additional shifts in block $k$; 1 otherwise \\
$\beta_{ii'k}$  & 0 If both employees $i$ and $i'$ have met or exceeded their minimum shift requirement and  \\
& employee $i$ is available to be scheduled for additional shifts in block $k$; 1 otherwise \\
$\gamma_{ii'k}$  & 0 If both employees $i$ and $i'$ have met or exceeded their minimum shift requirement and \\
& employee $i'$ is available to be scheduled for additional shifts in block $k$; 1 otherwise \\
$a_{ijk}$ & 1 if $\pi_{ik}$ = 0 and $X_{ijk}$ = 1 for shift $j$ in block $k$ for employee $i$; 0 otherwise \\
$b_{ii'jk}$ & 1 if $a_{ijk}$ = 1 and $\theta_{i'k}$ = 0 for shift $j$ in block $k$ for employees $i$ and $i'$; 0 otherwise \\
$F^{Demand-}_{ijk}$ & 1 if $\sum_{i'=0}^{i-1} b_{ii'jk} < D_{jk}$ for shift $j$ in block $k$ for employee $i$; 0 otherwise \\
$l_{ii'k}$ & 1 if $\delta_{ik} \leq \delta_{i'k} + 1$ for block $k$ for employees $i$ and $i'$; 0 otherwise \\
$t_{ii'k}$ & 1 if $\delta_{i'k} \leq \delta_{ik}$ for block $k$ for employees $i$ and $i'$; 0 otherwise \\
$h_{ii'jk}$ & 1 if $l_{ii'k}$ = 1 and $X_{ijk}$ = 1 for shift $j$ in block $k$ for employees $i$ and $i'$; 0 otherwise \\
$u_{ii'jk}$ & 1 if $t_{ii'k}$ = 1 and $X_{i'jk}$ = 1 for shift $j$ in block $k$ for employees $i$ and $i'$; 0 otherwise \\
$p_{ii'jk}$ & 1 if $h_{ii'jk}$ = 1 and $\theta_{i'k}$ = 1 for shift $j$ in block $k$ for employees $i$ and $i'$; 0 otherwise \\
$v_{ii'jk}$ & 1 if $u_{ii'jk}$ = 1 and $\theta_{ik}$ = 1 for shift $j$ in block $k$ for employees $i$ and $i'$; 0 otherwise \\
$F^{Demand+}_{ijk}$ & 1 if $\sum_{i'=0}^{i-1} n_{i'ijk} + \sum_{i'=i+1}^{n} v_{ii'jk} < D_{jk}$ for shift $j$ in block $k$ for employee $i$; 0 otherwise \\
\bottomrule

\end{longtable}
\end{center}

\setcounter{equation}{0}

\subsection{Objective Functions}
The integer program is executed in two separate stages, each has a different objective function.

\subsubsection*{Stage 1.}
In the first stage, we find the maximum demand that can be met with the availability provided by nurses: 
\begin{equation}
\nonumber
\min \ \sum_{j \in \Jset} \sum_{k \in \Kset} S_{jk} 
\end{equation}
The total demand subtracted by the optimal objective value of the first phase is equal to $D^{*}$.

\subsubsection*{Stage 2.}
In the second stage, we maximize nurse preferences while fulfilling the maximum demand that can be met with the availability provided by nurses, $D^{*}$. So, the objective is as follows:
\begin{equation}
\nonumber
\max \ \sum_{i \in \Iset} \sum_{j \in \Jset} \sum_{k \in \Kset} r_{ijk} X_{ijk} 
\end{equation}

\setcounter{equation}{0}

\subsection{Constraints}
In this section, we provide all the model constraints whose explanations can be found in Section \ref{subsec:ConstDesc}. For brevity, we omit the respective domains $\Iset, \Jset, \Kset$ of the $i,j,k$ indices in the forall parts of the constraints. We use $\mathbb{I}_{(\cdot)}$ to denote the indicator function, which returns 1 if the provided logical condition as the argument in the subscript is true, and 0 otherwise.

\subsubsection*{General Scheduling Constraints.}
\begin{alignat}{3}
& d_{jk} - S_{jk} - \sum_{i \in \Iset} X_{ijk} = 0 \quad && \forall j,k  \label{eq:1}\\
&X_{ijk} \leq y_{ijk} && \forall i,j,k  \label{eq:2}\\
&\sum_{j'=j-2}^{j} \left( \mathbb{I}_{(j' \geq 1)}  X_{i,j',k} + \mathbb{I}_{(j' \leq 0)} X_{i,\jind+j',k-1} \right) \leq 1 && \forall i,j,k \qquad  \label{eq:3} \\*[0.1cm]
& F^{available}_{ijk} = y_{ijk} && \forall i,j,k \label{eq:4} \\*[0.1cm]
&\sum_{j' \in \Jset} X_{ij'k} \leq g^{max} - F^{maxout}_{ijk} && \forall i,j,k  \label{eq:5} \\
&\sum_{j' \in \Jset} X_{ij'k} \geq g^{max}  (1- F^{maxout}_{ijk}) && \forall i,j,k  \label{eq:6} \\*[0.2cm]
&\!\!\!\sum_{j' = j-2}^{j+2}\left( \mathbb{I}_{(j' \in \Jset)}  X_{i j' k} + \mathbb{I}_{(j' \leq 0)} X_{i,\jind+j',k-1} + \mathbb{I}_{(j' \geq \jind+1)}  X_{i,j'-\jind,k+1} \right) \leq 2 (1  - F^{backtoback}_{ijk}) && \forall i,j,k \label{eq:7} \\
&\!\!\!\sum_{j' = j-2}^{j+2} \left( \mathbb{I}_{(j' \in \Jset)}  X_{i j' k} + \mathbb{I}_{(j' \leq 0)}  X_{i,\jind+j',k-1} + \mathbb{I}_{(j' \geq \jind+1)}  X_{i,j'-\jind,k+1} \right) \geq 1 - F^{backtoback}_{ijk} \ \ \ && \forall i,j,k \label{eq:8} \\
& F^{weekend}_{ijk} = 1 && \forall i,k, j \notin \bigcup_{m \in \Mset} \Lset_m \label{eq:9} \\
& \sum_{m' \in \Mset} \sum_{j' \in \Lset_{m'}} \sum_{k' \in \Kset} X_{i j'k'} \leq s \cdot r - w - F^{weekend}_{ijk} \quad && \forall i, k, j \in \bigcup_{m \in \Mset} \Lset_m 
\label{eq:10} \\
& \sum_{m' \in \Mset} \sum_{j' \in \Lset_{m'}} \sum_{k' \in \Kset} X_{i j'k'} \geq (s \cdot r - w)  (1 - F^{weekend}_{ijk}) \quad && \forall i, k, j \in \bigcup_{m \in \Mset} \Lset_m 
\label{eq:11} \\
& \sum_{e \in \Eset} (1-F^{e}_{ijk}) \geq 1-F_{ijk} && \forall i,j,k \label{eq:12} \\
& \sum_{e \in \Eset} (1-F^{e}_{ijk}) \leq (1-F_{ijk})  |\Eset| && \forall i,j,k \label{eq:13} \\*[0.3cm]
& \sum_{j \in \Jset} F_{ijk} \geq 1-M_{ik} && \forall i,k \label{eq:14} \\
& F_{ijk} \leq 1- M_{ik} && \forall i,j,k \label{eq:15}
\end{alignat}

\subsubsection*{Stage Two Additional General Scheduling Constraint.}
\begin{alignat}{3}
& \sum_{j \in \Jset} \sum_{k \in \Kset} S_{jk} \leq \  & \sum_{j \in \Jset} \sum_{k \in \Kset} D_{jk} - D^{*} &&  \label{eq:16}
\end{alignat}

\subsubsection*{Approximate Armstrong Constraints Set}
\begin{alignat}{3}
&F^{demand}_{1jk} \leq d_{jk}  && \forall j,k \label{eq:17} \\
&d_{jk}  (F^{demand}_{1jk} - 1) \geq 0  && \forall j,k \label{eq:18} \\
&\sum_{i' = 1}^{i-1} X_{i' jk} \leq d_{jk} - F^{demand}_{ijk} && \forall i\neq1, j,k \label{eq:19} \\
&\sum_{i' = 1}^{i-1} X_{i' jk} \geq d_{jk}  (1 - F^{demand}_{ijk}) && \forall i\neq 1,j,k \label{eq:20} \\
& \sigma_{ik} \geq \sigma_{(i+1)k} && \forall i \neq n, k  \label{eq:21} \\
& \theta_{ik} + M_{ik} \geq \sigma_{ik} && \forall i,k \label{eq:22} \\
& \sum_{j \in \Jset} X_{ijk} \geq g_{ik}  \theta_{ik} && \forall i,k \label{eq:23} \\
&  \sum_{j \in \Jset} X_{ijk} \leq g_{ik} \sigma_{ik} + (g_{ik} - 1)  (1-\sigma_{ik}) \quad  && \forall i,k \label{eq:24} 
\end{alignat}

\subsubsection*{Exact Armstrong Constraint Set}

\begin{alignat}{3}
& \sum_{j \in \Jset} X_{ijk} - g_{ik} = \delta_{ik} && \forall i,k \label{eq:25} \\
& \delta_{ik} \geq -g_{ik}(1 - \theta_{ik}) && \forall i,k \label{eq:26} \\
& \delta_{ik} \leq  (10 - g_{ik})\theta_{ik}  - 1 && \forall i,k \label{eq:27} \\
& \alpha_{ik} \leq M_{ik} + \theta_{ik} && \forall i,k \label{eq:28} \\
& \delta_{i'k} + g_{i'k} \leq \alpha_{ik} g^{max} && \forall i \neq n , i' \geq i+1 ,k \label{eq:29} \\
& \delta_{ik} \leq (10 - g_{ik}) \alpha_{i'k} && \forall i \neq n , i'  \geq i+1 ,k \label{eq:30} \\
& \beta_{ii'k} \leq (1 - \theta_{ik}) + (1 - \theta_{i'k}) + M_{ik} && \forall  i \neq n , i' \geq i+1 ,k \label{eq:31} \\
& \gamma_{ii'k} \leq (1 - \theta_{ik}) + (1 - \theta_{i'k}) + M_{i'k} && \forall  i \neq n , i' \geq i+1 ,k \label{eq:32} \\
& (\delta_{ik} - 1) - \delta_{i'k} \leq \gamma_{ii'k} (10 - g_{ik} + g_{i'k} - 1) && \forall i \neq n , i' \geq i+1 ,k \label{eq:33} \\
& \delta_{i'k} - \delta_{ik} \leq \beta_{ii'k} (10 - g_{i'k} + g_{ik}) && \forall i \neq n , i' \geq i+1 ,k \label{eq:34} \\
& \delta_{ik} \geq 1 - (g_{ik} + 1) (1-\pi_{ik}) && \forall i,k \label{eq:35} \\
& \delta_{ik} \leq  (10 - g_{ik}) \pi_{ik} && \forall i,k \label{eq:36} \\
& a_{ijk} \leq 1 - \pi_{ik} && \forall i,j,k \label{eq:37} \\
& a_{ijk} \leq X_{ijk} && \forall i,j,k \label{eq:38} \\
& a_{ijk} \geq X_{ijk} - \pi_{ik} && \forall i,j,k \label{eq:39}\\
& b_{ii'jk} \leq 1 - \theta_{i'k} && \forall i \neq n , i' \geq i+1 ,j,k \label{eq:40} \\
& b_{ii'jk} \leq a_{ijk} && \forall i \neq n , i' \geq i+1 ,j,k \label{eq:41} \\
& b_{ii'jk} \geq a_{ijk} - \theta_{i'k} && \forall i \neq n , i' \geq i+1 ,j,k \label{eq:42} \\
& \sum_{i'=0}^{i-1} b_{i'ijk} \leq D_{jk} - F_{ijk}^{Demand-} && \forall i,j,k \label{eq:43} \\
& \sum_{i'=0}^{i-1} b_{i'ijk} \geq D_{jk}  (1 - F_{ijk}^{Demand-}) && \forall i,j,k \label{eq:44} \\
& \delta_{ik} - (10 - g_{ik} + g_{i'k} + 1)(1-l_{ii'k}) \leq \delta_{i'k} + 1 && \forall i \neq n , i' \geq i+1 ,k \label{eq:45}\\
& \delta_{ik} + (10 - g_{i'k} + g_{ik} + 2) l_{ii'k} \geq \delta_{i'k} + 2 && \forall i \neq n , i' \geq i+1 ,k \label{eq:46}\\
& h_{ii'jk} \leq l_{ii'k} && \forall i \neq n , i' \geq i+1 ,j,k \label{eq:47} \\
& h_{ii'jk} \leq X_{ijk} && \forall i \neq n , i' \geq i+1 ,j,k \label{eq:48} \\
& h_{ii'jk} \geq X_{ijk} - (1 - l_{ii'k}) && \forall i \neq n , i' \geq i+1 ,j,k \label{eq:49} \\
& p_{ii'jk} \leq h_{ii'jk} && \forall i \neq n , i' \geq i+1 ,j,k \label{eq:50} \\
& p_{ii'jk} \leq \theta_{i'k} && \forall i \neq n , i' \geq i+1 ,j,k \label{eq:51} \\
& p_{ii'jk} \geq \theta_{i'k} - (1 - h_{ii'jk}) && \forall i \neq n , i' \geq i+1 ,j,k \label{eq:52} \\
& \delta_{i'k} - (10 - g_{i'k} + g_{ik})(1-t_{ii'k}) \leq \delta_{ik} && \forall i \neq n , i' \geq i+1 ,k \label{eq:53}\\
& \delta_{i'k} + (10 - g_{ik} + g_{i'k} + 1) t_{ii'k} \geq \delta_{ik} + 1 && \forall i \neq n , i' \geq i+1 ,k \label{eq:54}\\
& u_{ii'jk} \leq t_{ii'k} && \forall i \neq n , i' \geq i+1 ,j,k \label{eq:55} \\
& u_{ii'jk} \leq X_{i'jk} && \forall i \neq n , i' \geq i+1 ,j,k \label{eq:56} \\
& u_{ii'jk} \geq X_{i'jk} - (1 - t_{ii'k}) && \forall i \neq n , i' \geq i+1 ,j,k \label{eq:57} \\
& v_{ii'jk} \leq u_{ii'jk} && \forall i \neq n , i' \geq i+1 ,j,k \label{eq:58} \\
& v_{ii'jk} \leq \theta_{ik} && \forall i \neq n , i' \geq i+1 ,j,k \label{eq:59} \\
& v_{ii'jk} \geq \theta_{ik} - (1 - u_{ii'jk}) && \forall i \neq n , i' \geq i+1 ,j,k \label{eq:60} \\
& \sum_{i'=0}^{i-1}n_{i'ijk} + \sum_{i'=i+1}^{I}v_{ii'jk} \leq D_{jk} - F_{ijk}^{Demand+} && \forall i,j,k \label{eq:61} \\
& \sum_{i'=0}^{i-1}n_{i'ijk} + \sum_{i'=i+1}^{I}v_{ii'jk} \geq D_{jk} (1 - F_{ijk}^{Demand+}) \quad && \forall i,j,k \label{eq:62} \\
&  F_{ijk}^{Demand} \leq F_{ijk}^{Demand-} + \theta_{ik} && \forall i,j,k \label{eq:63} \\
&  F_{ijk}^{Demand} \leq F_{ijk}^{Demand+} + (1 - \theta_{ik}) && \forall i,j,k \label{eq:64} \\
&  F_{ijk}^{Demand} \geq F_{ijk}^{Demand+} + (\theta_{ik} - 1) && \forall i,j,k \label{eq:65} \\
&  F_{ijk}^{Demand} \geq F_{ijk}^{Demand-} - \theta_{ik} && \forall i,j,k  
\label{eq:66} 
\end{alignat}

\subsection{Constraint Descriptions}
\label{subsec:ConstDesc}
The following is a description for each constraint of the General and Armstrong scheduling constraint sets. 

\subsubsection*{General Scheduling Constraint Descriptions.}
\begin{enumerate}
    \item[{\eqref{eq:1}}]  The total supply (total shifts assigned plus unfilled demand) must equal to the demand for any given shift.
    \item[{\eqref{eq:2}}] A nurse cannot be scheduled if they are not available.
    \item[{\eqref{eq:3}}] A shift cannot be scheduled within two shifts of another scheduled shift for the same nurse. Note that for a given block $k$, if $j' = 0$ and $j' = -1$ then the shift $j'$ preceding $j$ corresponds to the last (i.e.,  $q^{\text{th}}$) and the second to the last (i.e., $(q-1)^{\text{th}}$) shift of the block $k-1$, respectively, thus the shift index in the second term in parenthesis is $q + j'$. For $k=1$, those previous shifts are considered as the status parameters of the nurse for the last two shifts of the previous planning horizon.  
    \item[{\eqref{eq:4}}] If a nurse is available for a given shift set $F^{available}_{ijk}$ to 1 for that shift for that nurse otherwise set to 0. ($F^{available}$ variables are introduced for notational consistency with the other exception related variables.)   
    \item[{\eqref{eq:5}}] A nurse cannot be assigned more than $g^{max}$ shifts but if they are assigned exactly $g^{max}$ shifts in a given block then set the $F^{maxout}_{ijk}$ variable to 0 for all shifts in the given block for that nurse.
    \item[{\eqref{eq:6}}] Given a shift and a nurse, if the nurse is assigned less than $g^{max}$ shifts for the block in which the shift falls, then set $F^{maxout}_{ijk}$ variable to 1.
    \item[{\eqref{eq:7}}] Given a shift and a nurse, if the nurse has been scheduled for at least one shift within two shifts of the given shift, the left-hand side of the constraint amounting to at least 1 out of 5 consecutive shifts considered, then set the $F^{backtoback}_{ijk}$ variable to 0. (Note: The indicator variables account for back-to-back shifts which span between two-week blocks in the six-week horizon.)
    \item[{\eqref{eq:8}}] Given a shift and a nurse, if a nurse has not been scheduled for a shift within two shifts of the given shift, set the $F^{backtoback}_{ijk}$ variable to 1.
    \item[{\eqref{eq:9}}] For all shifts which do not fall on a weekend, set the $F^{weekend}_{ijk}$ variable to 1 for all nurses. 
    \item[{\eqref{eq:10}}] Given a weekend shift and a nurse, if the nurse has been scheduled for the maximum number of weekend shifts allowable over the six-week horizon, $s \cdot r - w$ (i.e., the total number of weekend shifts minus the minimum number of required weekend days off), set the $F^{weekend}_{ijk}$ variable to 0 for that shift.
    \item[{\eqref{eq:11}}] Given a weekend shift and a nurse, if the nurse has not yet been scheduled for the maximum number of weekend shifts allowable over the six-week horizon, set the $F^{weekend}_{ijk}$ variable to 1.
    \item[{\eqref{eq:12}}] For a given nurse and a given shift, if all $F_{ijk}$ exception variables ($F^{available}_{ijk}$, $F^{demand}_{ijk}$, $F^{backtoback}_{ijk}$, $F^{maxout}_{ijk}$, $F^{weekend}_{ijk}$) for that nurse and shift are equal to 1, set $F_{ijk}$ to 1. 
    \item[{\eqref{eq:13}}] For a given nurse and a given shift, if any $F_{ijk}$ exception variables ($F^{available}_{ijk}$, $F^{demand}_{ijk}$, $F^{backtoback}_{ijk}$, $F^{maxout}_{ijk}$, $F^{weekend}_{ijk}$) for that nurse and shift are equal to 0, set $F_{ijk}$ to 0. 
    \item[{\eqref{eq:14}}] For a given nurse, if all $F_{ijk}$ variables in a given block are equal to 0, then the nurse cannot get any more shifts, thus set $M_{ik}$ equal to 1 for that nurse and block. 
    \item[{\eqref{eq:15}}] If the $F_{ijk}$ variable for a given nurse and shift is equal to 1, set $M_{ik}$ equal to 0 for that nurse for the block which the shift belongs to. 
    \item[{\eqref{eq:16}}] The total unfilled demand across the six-week scheduling horizon for the second stage should be less than or equal to the total demand over the horizon minus the maximum demand that can be filled over the horizon with the availability provided by nurses. This constraint ensures that demand is not left unfilled in order to improve the preference of shifts assigned.
\end{enumerate}
    
\subsubsection*{Approximate Armstrong Constraint Set Descriptions.}    
    \begin{enumerate}
     \item[{\eqref{eq:17}}] If the demand is 0 for a given shift, set the $F^{demand}_{ijk}$ variable to 0 for the highest seniority nurse.
    \item[{\eqref{eq:18}}] If the demand is greater than or equal to 1 for a given shift, set the $F^{demand}_{ijk}$ variable to 1 for the highest seniority nurse.
    \item[{\eqref{eq:19}}] Given a nurse who is not the highest seniority nurse, if the total number of shifts assigned to all nurses of higher seniority than the given nurse is equal to the demand for a given shift, then set the $F^{demand}_{ijk}$ variable to 0.
    \item[{\eqref{eq:20}}] Given a nurse who is not the highest seniority, if the total number of shifts assigned to all nurses of higher seniority than the given nurse is less than the demand for a given shift, then set the $F^{demand}_{ijk}$ variable to 1.
    \item[{\eqref{eq:21}}] Ensure that higher seniority nurses are being assigned their minimum shift requirement or hit an exception before lower seniority nurses.
    \item[{\eqref{eq:22}}] If $\theta_{ik}$ and $M_{ik}$ are both 0 for a given nurse and block, $\sigma_{ik}$ must also equal to 0 for that nurse in that block, indicating that the nurse has not hit their minimum shift requirement nor an exception. 
    \item[{\eqref{eq:23}}] If a nurse is assigned their minimum shift requirement in a given block, their $\theta_{ik}$ variable must be set to 1 for that block. 
    \item[{\eqref{eq:24}}] If $\sigma_{ik}$ is set to 1 for a given nurse and block then the nurse can be assigned at most their minimum shift requirement, otherwise if $\sigma_{ik}$ is set to 0, the nurse should be assigned strictly less than their minimum shift requirement.  
    (Note that if the second part of the right-hand-side is removed, the approximation becomes exact for the case where total demand is less than or equal to total minimum shift requirement).
\end{enumerate}

\subsubsection*{Exact Armstrong Constraint Set Descriptions.} 
    \begin{enumerate}
    \item[{\eqref{eq:25}}] For a given nurse, set $\delta_{ik}$ as the difference between the shifts assigned to the nurse in a given block minus their minimum shift requirement for that block.
    \item[{\eqref{eq:26}}] If $\delta_{ik}$ is less than 0, enforce $\theta_{ik}$ to be 0.
    \item[{\eqref{eq:27}}] If $\delta_{ik}$ is greater than or equal to 0, enforce $\theta_{ik}$ to be 1.
    \item[{\eqref{eq:28}}] If $\theta_{ik}$ and $M_{ik}$ are equal to 0 for a given nurse and block, $\alpha_{ik}$ must also be equal to 0. 
    \item[{\eqref{eq:29}}] If $\alpha_{ik}$ is equal to 0 for a given nurse and block, each nurse with less seniority than them should have a $\delta_{ik}$ equal to the negative of their minimum shift requirement.
    \item[{\eqref{eq:30}}] If $\alpha_{ik}$ is equal to 0 for a given nurse and block, each nurse with more seniority than them should not have an Armstrong delta which exceeds 0. 
    \item[{\eqref{eq:31}}] If both nurse $i$ and $i'$ (where $i$ $\leq$ $i'$ + 1) have met or exceeded their minimum shift requirement ($\theta_{ik}$ = 1, $\theta_{i'k}$ = 1) and nurse $i$ can still be scheduled without hitting an exception ($M_{ik}$ = 0) then set $\beta_{ii'k}$ = 0.
    \item[{\eqref{eq:32}}] If both nurse $i$ and $i'$ (where $i$ $\leq$ $i'$ + 1) have met or exceeded their minimum shift requirement ($\theta_{ik}$ = 1, $\theta_{i'k}$ = 1) and nurse $i'$ can still be scheduled without hitting an exception ($M_{i'k}$ = 0) then set $\gamma_{ii'k}$ = 0.
    \item[{\eqref{eq:33}}] If $\gamma_{ii'k}$ is equal to 0, then $\delta_{ik}$ should be at most one greater than $\delta_{i'k}$.
    \item[{\eqref{eq:34}}] If $\beta_{ii'k}$ is equal to 0, then $\delta_{ik}$ should be greater than $\delta_{i'k}$.
    \item[{\eqref{eq:35}}] If $\delta_{ik}$ is less than or equal to 0, enforce $\pi_{ik}$ to be 0.
    \item[{\eqref{eq:36}}] If $\delta_{ik}$ is greater than 0, enforce $\pi_{ik}$ to be 1.
    \item[{\eqref{eq:37}}] If $\pi_{ik}$ is 1, enforce $a_{ijk}$ to be 0.
    \item[{\eqref{eq:38}}] If $X_{ijk}$ is 0, enforce $a_{ijk}$ to be 0.
    \item[{\eqref{eq:39}}] If $X_{ijk}$ is 1 and $\pi_{ik}$ is 0, enforce $a_{ijk}$ to be 1.
    \item[{\eqref{eq:40}}] For nurses $i$ and $i'$ (where $i$ $\leq$ $i'$ + 1), if $\theta_{i'k}$ is 1, enforce $b_{ii'jk}$ to be 0.
    \item[{\eqref{eq:41}}] For nurses $i$ and $i'$ (where $i$ $\leq$ $i'$ + 1), if $a_{ijk}$ is 0, enforce $b_{ii'jk}$ to be 0.
    \item[{\eqref{eq:42}}] If $a_{ijk}$ is 1 and $\theta_{i'k}$ is 0, enforce $b_{ii'jk}$ to be 1.
    \item[{\eqref{eq:43}}] For nurse $i$ and shift j in block k, if the sum of all $b_{i'ijk}$ variables for all nurses $i'$ who are more senior to nurse $i$ is equal to the demand for that shift, set $F_{ijk}^{Demand-}$ to 0.
    \item[{\eqref{eq:44}}] For nurse $i$ and shift j in block k, if the sum of all $b_{i'ijk}$ variables for all nurses $i'$ who are more senior to nurse $i$ is less than the demand for that shift, set $F_{ijk}^{Demand-}$ to 1.
    \item[{\eqref{eq:45}}] For nurses $i$ and $i'$ (where $i$ $\leq$ $i'$ + 1), if $\delta_{ik}$ is greater than $\delta_{i'k}$ + 1 set $l_{i'ik}$ to 0. 
    \item[{\eqref{eq:46}}] For nurses $i$ and $i'$ (where $i$ $\leq$ $i'$ + 1), if $\delta_{ik}$ is less than or equal to $\delta_{i'k}$ + 1 set $l_{i'ik}$ to 1. 
    \item[{\eqref{eq:47}}] For nurses $i$ and $i'$ (where $i$ $\leq$ $i'$ + 1), if $l_{ii'k}$ is 0, enforce $h_{ii'jk}$ to be 0.
    \item[{\eqref{eq:48}}] For nurses $i$ and $i'$ (where $i$ $\leq$ $i'$ + 1), if $X_{ijk}$ is 0, enforce $h_{ii'jk}$ to be 0.
    \item[{\eqref{eq:49}}] If $X_{ijk}$ is 1 and $l_{ii'k}$ is 1, enforce $h_{ii'jk}$ to be 1.
    \item[{\eqref{eq:50}}] For nurses $i$ and $i'$ (where $i$ $\leq$ $i' + 1$), if $h_{ii'jk}$ is 0, enforce $p_{ii'jk}$ to be 0.
    \item[{\eqref{eq:51}}] For nurses $i$ and $i'$ (where $i$ $\leq$ $i'$ + 1), if $\theta_{i'k}$ is 0, enforce $p_{ii'jk}$ to be 0.
    \item[{\eqref{eq:52}}] If $\theta_{i'k}$ is 1 and $h_{ii'jk}$ is 1, enforce $h_{ii'jk}$ to be 1.
    \item[{\eqref{eq:53}}] For nurses $i$ and $i'$ (where $i$ $\leq$ $i'$ + 1), if $\delta_{i'k}$ is greater than $\delta_{ik}$ set $t_{i'ik}$ to 0.
    \item[{\eqref{eq:54}}] For nurses $i$ and $i'$ (where $i$ $\leq$ $i'$ + 1), if $\delta_{i'k}$ is less than or equal to $\delta_{ik}$ set $t_{i'ik}$ to 1. 
    \item[{\eqref{eq:55}}] For nurses $i$ and $i'$ (where $i$ $\leq$ $i'$ + 1), if $t_{ii'k}$ is 0, enforce $u_{ii'jk}$ to be 0.
    \item[{\eqref{eq:56}}] For nurses $i$ and $i'$ (where $i$ $\leq$ $i'$ + 1), if $X_{i'jk}$ is 0, enforce $u_{ii'jk}$ to be 0.
    \item[{\eqref{eq:57}}] If $X_{i'jk}$ is 1 and $t_{ii'k}$ is 1, enforce $u_{ii'jk}$ to be 1.
    \item[{\eqref{eq:58}}] For nurses $i$ and $i'$ (where $i$ $\leq$ $i'$ + 1), if $u_{ii'jk}$ is 0, enforce $v_{ii'jk}$ to be 0.
    \item[{\eqref{eq:59}}] For nurses $i$ and $i'$ (where $i$ $\leq$ $i'$ + 1), if $\theta_{ik}$ is 0, enforce $v_{ii'jk}$ to be 0.
    \item[{\eqref{eq:60}}] If $\theta_{ik}$ is 1 and $u_{ii'jk}$ is 1, enforce $v_{ii'jk}$ to be 1.
    \item[{\eqref{eq:61}}] For an nurse $i$ and shift j in block $k$, if the sum of all $n_{i'ijk}$ variables for all nurses $i'$ who are more senior to nurse $i$ plus the sum of all $v_{i'ijk}$ variables for all nurses $i'$ who are less senior to nurse $i$ is equal to the demand for that shift, set $F_{ijk}^{Demand+}$ to 0.
    \item[{\eqref{eq:62}}] For an nurse $i$ and shift j in block $k$, if the sum of all $n_{i'ijk}$ variables for all nurses $i'$ who are more senior to nurse $i$ plus the sum of all $v_{i'ijk}$ variables for all nurses $i'$ who are less senior to nurse $i$ is less than the demand for that shift, set $F_{ijk}^{Demand+}$ to 1.
    \item[{\eqref{eq:63}}] For a nurse $i$ and shift $j$ in block $k$, if $F_{ijk}^{Demand-}$ is 0 and $\theta_{ik}$ is 0, set $F_{ijk}^{Demand}$ to 0.
    \item[{\eqref{eq:64}}] For a nurse $i$ and shift $j$ in block $k$, if $F_{ijk}^{Demand+}$ is 0 and $\theta_{ik}$ is 1, set $F_{ijk}^{Demand}$ to 0. 
    \item[{\eqref{eq:65}}] For a nurse $i$ and shift $j$ in block $k$, if $F_{ijk}^{Demand+}$ is 1 and $\theta_{ik}$ is 1, set $F_{ijk}^{Demand}$ to 1.
    \item[{\eqref{eq:66}}] For a nurse $i$ and shift $j$ in block $k$, if $F_{ijk}^{Demand-}$ is 1 and $\theta_{ik}$ is 0, set $F_{ijk}^{Demand}$ to 1.
\end{enumerate}

\subsection{IP Descriptions} \label{subsec:MIPDesc}
The approximate and exact models both use a similar two-stage approach to solve the problem each with a different set of constraints. The main decision variables for both IPs are the nurse schedules ($X_{ijk}$) and unfilled demand ($S_{jk}$). They include a large set of binary indicator variables to represent certain logical conditions in a linear fashion (via the so-called big-M constraints). In the first phase of both models the outputted schedule is insignificant, as the main objective of this phase is to determine the minimum amount of unfilled demand that be scheduled while meeting all of the general scheduling constraints and Armstrong Seniority constraints. The general scheduling constraints are enforced in both IPs using the $F_{ijk}$ exception variables. These variables help to ensure that nurse availability is respected ($F_{ijk}^{available}$), no overtime is incurred ($F_{ijk}^{maxout}$), back-to-back shifts are not assigned ($F_{ijk}^{backtoback}$), nurses do not exceed a certain number of weekend shifts ($F_{ijk}^{weekend}$), and nurse demand is never exceeded ($F_{ijk}^{demand}$). If any one or more of these exception variables is activated (set to 0) for a given shift $j$ in a block $k$ and employee $i$, the employee cannot be assigned the shift, which is indicated by the $F_{ijk}$ variable being set to 0. If employee $i$ is unable to get more shifts in block $k$ than what is assigned in a solution, $M_{ik}$ variable is activated (set to 1).

There are two sets of Armstrong Seniority constraints, the first set is the approximate Armstrong constraint set which enforces an approximation of the Armstrong rule and must be utilized in tandem with the heuristic algorithm in order to ensure that the Armstrong seniority requirements are met. The second set is the exact Armstrong constraint set which is an exact formulation of the Armstrong seniority requirements. In the second phase of both approaches, the objective is to maximize the overall shift preferences of the nurses while meeting the maximum demand level that can be met found in the first phase, while satisfying the general scheduling and Armstrong Seniority constraints. Detailed constraint descriptions can be found in section \ref{subsec:ConstDesc}.

The approximate Armstrong constraint set is utilized in the IP formulation for the implemented approach. The main logic of this constraint set is to prioritize seniority while assigning each nurse as many shifts as possible up to their minimum shift requirement, but no more than their minimum shift requirement. Each nurse can only be assigned their minimum shift requirement if the nurse directly senior to them has been assigned their minimum shift requirement or if they cannot be scheduled for any additional shifts without breaking a general scheduling rule. This is accomplished by this $\sigma_{ik}$ variables; specifically if nurse $i$ has not hit their minimum shift requirement in block $k$ nor an exception, $\sigma_{ik}$ is set to 0, which in turns sets $\sigma_{\cdot k}$ variables for all the lower seniority nurses, thus blocks them to receive any shifts at all. This constraint set is an approximation for two reasons: (1) It does not allow nurses to be assigned more shifts than their minimum shift requirement and (2) it allows a nurse of lower seniority to be assigned up to one less than their minimum shift requirement even if the nurse(s) senior to them have not been assigned their minimum shift requirement. Due to these factors, the resulting schedule may not always abide by the Armstrong scheduling rules, therefore the heuristic algorithm is used to ensure a feasible schedule. Note that if the second part of the right-hand side of constraint \eqref{eq:24} is removed, this approximation becomes exact for the case in which total demand is less than or equal to total minimum shift requirements. However, it might lead to a computationally challenging model (which is indeed observed in our numerical experiments).

The exact Armstrong constraint set is implemented in the Exact IP model. This set ensures the Armstrong seniority requirements are met and an optimal feasible solution is provided for any combination of availability and demand provided. The main logic of this constraint set revolves around the Armstrong delta ($\delta_{ik}$) variables and a large set of layered binary indicator variables which together are used to activate the $F_{ijk}^{Demand}$ variable for a given employee, block, and shift.
Put simply, the $F_{ijk}^{Demand}$ variable indicates whether or not an employee is available to be scheduled for a shift based upon whether or not the demand is filled according to the Armstrong scheduling rules.
The $F_{ijk}^{Demand}$ variable is activated by the $F_{ijk}^{Demand-}$ and $F_{ijk}^{Demand+}$ variables. The $F_{ijk}^{Demand-}$ variable is used to indicate whether an employee is available to be scheduled for a shift if they have been assigned less than their minimum shift requirement and is activated based upon the number of shifts assigned to employees of higher seniority who are also available for the shift. The $F_{ijk}^{Demand+}$ variable is used to indicate whether an employee is available to be scheduled for a shift if they have been assigned exactly or more than their minimum shift requirement and is activated based upon the number of shifts assigned to employees of higher and lower seniority who are also available for the shift. The $F_{ijk}^{Demand}$ variables (along with the other $F_{ijk}$ exception variables) are used to activate the $F_{ijk}$ variables, which are then used to activate the $M_{ik}$ variables. The $M_{ik}$ variable indicates if an employee has hit an exception and is unable to be scheduled for any additional shifts in a given block without breaking a general scheduling or Armstrong seniority requirement. The entire $F_{ijk}^{Demand}$ logic is needed due to the introduction of the minimum shift requirements and the Armstrong seniority rules. This variable helps keeping track of whether a shift can be assigned to a nurse based upon the set of all other assigned shifts in the schedule. More specifically, if  employee $i$ is eligible to receive shift $j$ of block $k$ considering all the exceptions other than the demand, but has not received the shift, then the other employees who are eligible to fill that shift are taken into consideration. If the nurse has not hit their minimum shift requirement for the block, then only their seniors are eligible to fill the shift. On the other hand, if the nurse $i$ has $\delta_{ik} \geq 0$, then a more senior nurse $i'$ is eligible to fill the shift if $\delta_{i'k} \leq \delta_{ik} + 1$, whereas a less senior nurse $i'$ is eligible if $\delta_{i'k} \leq \delta_{ik}$. The $F_{ijk}^{Demand}$ variable is correctly activated based on the status of those eligible nurses.  

\section{Heuristic Algorithm}\label{Apendix:Heuristic}
The heuristic, provided in Algorithm \ref{Alg1}, takes the schedule outputted by the IP model as an input. It begins by attempting to assign each unfilled shift to the most senior eligible nurse, if any, and updates the assignments, as well as the eligibility and exception variables (Steps 2-8). 
Once all of the unfilled demand that is able to be filled is assigned, the shift reassignment phase of the heuristic begins. In this phase of the algorithm (Steps 9-30), the schedule is gradually modified in order to satisfy the Armstrong requirements. 
This phase consists of a check and reassign routine which loops over all the shifts and for each shift considers all nurse pairs with one nurse assigned and the other not assigned to the shift, but eligible to be assigned.
The routine checks to see whether the nurse who is currently assigned the shift is the more senior or less senior nurse. If the shift is assigned to the more senior nurse, it checks whether the less senior nurse should be reassigned the shift. The less senior nurse will be reassigned the shift if either (1) the less senior nurse has nonnegative Armstrong delta, and the more senior nurse has at least two more shifts than the less senior nurse, or (2) the more senior nurse has a positive Armstrong delta and the less senior nurse has a negative Armstrong delta. If the shift is assigned to the less senior nurse, it checks whether the more senior nurse should be reassigned the shift. The more senior nurse will be reassigned the shift if (1) both nurses have a positive Armstrong delta and the less senior nurse has an Armstrong delta greater than the more senior nurse; (2) the less senior nurse has a positive Armstrong delta and the more senior nurse has an Armstrong delta of zero; or (3) the more senior nurse has a negative Armstrong delta. If any of these scenarios occur then one of the Armstrong rules is violated and the shift is reassigned to the other nurse. Once this procedure is executed for all assigned shifts, a single iteration is complete. The procedure is then repeated for a number of times set by the user (Iteration Limit). The outputs of the heuristic include the final schedule ($X_{ijk}$ variables), unfilled demand ($S_{jk}$ variables), and all exception status ($F_{ijk}$  variables). 

\begin{algorithm}
\footnotesize
\caption{Heuristic Algorithm}
\label{Alg1}
\begin{algorithmic} [1] 
\STATE \textbf{Inputs:} Schedule, Exception Variables, Unfilled Demand, Demand, Iteration Limit
\FOR{each block $k\in \Kset$}
\FOR{each shift $j\in \Jset$}
\FOR{each nurse $i =  1,\hdots,n$ }
    \IF{Not all shifts have been assigned and the nurse is able to be scheduled without breaking a general scheduling rule ($S_{jk}$ $>$ 0, $F_{ijk}$ = 1)}
    \STATE $X_{ijk}$ = 1 (Assign the shift to the nurse)
    \STATE $S_{jk} = S_{jk} - 1$
    \STATE Update $\delta_{ik}$ and $F_{ijk}$ variables
    \ENDIF
\ENDFOR
\ENDFOR
\ENDFOR
\STATE 
Set \ShiftSwaps \ = 1

\WHILE{\ShiftSwaps \ $>$ 0}
\STATE Set \ShiftSwaps \ = 0
\FOR{each block $k \in \Kset$}
\STATE Update $\delta_{ik}$ variables
\FOR{each shift $j \in \Jset$}
\IF{$D_{jk} >$ 0}
\FOR{each nurse $i^{'} = 1,\hdots,n - 1$}
\FOR{each nurse $i^{''} = i^{'} + 1,\hdots,n$}
\IF{$X_{i^{'}jk}$ = 1 $\And$ $F_{i^{''}jk}$ = 1}
\IF{$\delta_{i^{'}k}$ $>$ 0 $\And$ $\delta_{i^{''}k}$ $\geq$ 0}
\IF{$\delta_{i^{'}k}$ - $\delta_{i^{''}k}$ $>$ 1}
\STATE Set $X_{i^{'}jk}$ = 0, $X_{i^{''}jk}$ = 1, \ShiftSwaps \ = \ShiftSwaps \ + 1
\ENDIF
\ELSIF{$\delta_{i^{'}k}$ $>$ 0 $\And$ $\delta_{i^{''}k}$ $<$ 0}
\STATE Set $X_{i^{'}jk}$ = 0, $X_{i^{''}jk}$ = 1, \ShiftSwaps \ = \ShiftSwaps \ + 1
\ENDIF
\ENDIF
\IF{$X_{i^{''}jk}$ = 1 $\And$ $F_{i^{'}jk}$ = 1}
\IF{$\delta_{i^{''}k}$ $>$ 0 $\And$ $\delta_{i^{'}k}$ $>$ 0}
\IF{$\delta_{i^{''}k}$ - $\delta_{i^{'}k}$ $>$ 0}
\STATE Set $X_{i^{''}jk}$ = 0, $X_{i^{'}jk}$ = 1, \ShiftSwaps \ = \ShiftSwaps \ + 1
\ENDIF
\ELSIF{($\delta_{i^{''}k}$ $>$ 0 $\And$ $\delta_{i^{'}k}$ $=$ 0) OR ($\delta_{i^{'}k}$ $<$ 0)}
\STATE Set $X_{i^{''}jk}$ = 0, $X_{i^{'}jk}$ = 1, \ShiftSwaps \ = \ShiftSwaps \ + 1
\ENDIF
\ENDIF
\STATE Update $\delta_{ik}$ and $F_{ijk}$ variables
\ENDFOR
\ENDFOR
\ENDIF
\ENDFOR
\ENDFOR
\ENDWHILE
\STATE \textbf{Outputs:} Schedule, Unfilled Demand, Exception Variables
\end{algorithmic}
\end{algorithm}

The heuristic stopping criteria is based upon how many swaps occur within one iteration of the algorithm. If no swaps occur, the algorithm stops since all general scheduling and Armstrong scheduling criteria were met for all shifts checked within that iteration. The check and reassign routine is guaranteed to terminate after a finite number of iterations as an individual shift (one unit of demand) cannot be reassigned back to a given nurse once it has been taken away from them. In our experiments, the algorithm terminated after at most three iterations and returned a feasible schedule. Below we provide a formal argument for the above claim as well as the feasibility of the output schedule. 

\textbf{The algorithm terminates after a finite number of iterations.} If a shift is re-assigned to an employee of lower seniority according to lines 18-21, both the $\delta_{i'k}$ and $\delta_{i''k}$ variables must be greater than 0 after the reassignment, with the $\delta_{i''k}$ variable no greater than  $\delta_{i'k}$. If a shift is reassigned to an employee of lower seniority according to lines 22-23, the $\delta_{i'k}$ variable must be greater than or equal to zero and the $\delta_{i''k}$ variable must less than or equal to zero after the reassignment. In both cases, this shift can not be reassigned back to an employee of higher seniority via lines 25-27 or lines 28-29 as the following conditions must be false: $\delta_{i''k}$ $>$ 0 \& $\delta_{i'k}$ $>$ 0 \& $\delta_{i''k}$ - $\delta_{i'k}$ $>$ 0, $\delta_{i''k}$ $>$ 0 \& $\delta_{i'k}$ $=$ 0, and $\delta_{i'k}$ $<$ 0. If a shift is reassigned to an employee of higher seniority according to lines 25-27, the $\delta_{i'k}$ variable must be greater than or equal to two and the $\delta_{i''k}$ variable must be greater than or equal to one after the reassignment, with the $\delta_{i'k}$ variable at most one more than $\delta_{i''k}$. If a shift is reassigned to an employee of higher seniority according to lines 28-29, the $\delta_{i'k}$ variable must be less than or equal to one and the $\delta_{i''k}$ variable must be greater than or equal to zero after the reassignment. This shift can not be reassigned back to an employee of lower seniority via lines 18-21 or lines 22-23 as the following conditions must be false: $\delta_{i'k}$ $>$ 0 \& $\delta_{i''k}$ $\geq$ 0 \& $\delta_{i'k}$ - $\delta_{i''k}$ $>$ 1, and $\delta_{i'k}$ $>$ 0 \& $\delta_{i''k}$ $<$ 0. It can be shown that these conditions hold not only for individual reassignments but also for a chain of reassignments. Therefore, there is only a finite number of feasible reassignments that can occur. This guarantees that the heuristic terminates after a finite number of iterations. 

\textbf{The output schedule is feasible.} The outputted schedule is feasible if it abides by the general scheduling requirements and as long as no individual scheduled shift can be reassigned within the same shift to a more or less senior nurse without breaking an Armstrong seniority rule. Therefore, the implemented check and reassign routine guarantees feasibility as it will reassign all shifts that do not abide by any of the Armstrong seniority requirements. More specifically, Rule 1A is enforced by the reassignment made in lines 28-29 (the latter portion of the OR statement); Rule 1B is enforced by the checks made in lines 22-23; and Rule 2 is enforced by the checks made in lines 19-21, 25-27, and 28-29 (the first portion of the OR statement). All feasible individual reassignments must be completed by the algorithm before it can terminate. The algorithm completes shift reassignments within a shift and never between shifts, and is indifferent to the different preference scores and amount of unfilled demand in a candidate schedule. This means while the outputted schedule is guaranteed to abide by the Armstrong seniority rules, it is not guaranteed to be optimal with respect to the number of unfilled demand remaining or the preferences of the nurses.

\newpage

\begin{landscape}

\section{Schedule Output} \label{Apendix:ScheduleOutput}

\begin{figure}[htbp] 
\centering
\includegraphics[scale = 1.35,angle=0]{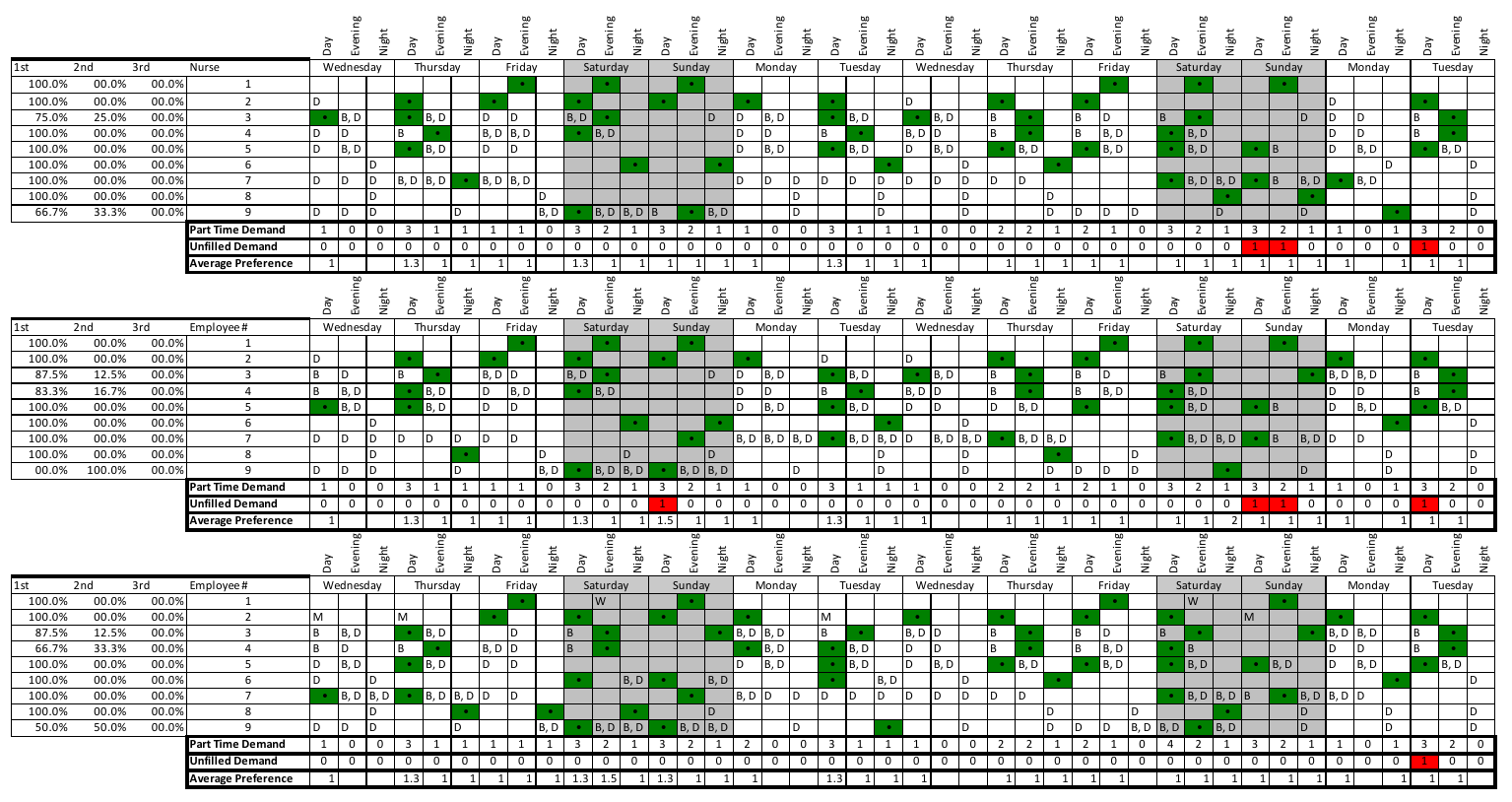}
\caption{City of Toronto LTCH\&S Division scheduling tool schedule output for a full six week planning horizon}
\label{fig:ExampleSchedule_full}
\end{figure}

\end{landscape}

\end{APPENDICES}

\end{document}